\documentclass[aps,prx,twocolumn,showkeys,superscriptaddress,nobibnotes]{revtex4-2}

\usepackage{placeins}                   
\usepackage[artemisia]{textgreek}       
\usepackage{amssymb}                    
\usepackage{amsmath} 				    
\usepackage{latexsym}                   
\usepackage{nicefrac}                   

\usepackage{graphicx}                   
\usepackage{epsfig}                     
\usepackage{epstopdf}                   
\usepackage{dcolumn}                    

\usepackage{lipsum}                     
\usepackage{color}                      

\usepackage{subfigure}                  
\usepackage{blindtext}                  
\usepackage[ulem=normalem]{changes}     
\usepackage{float}                      
\usepackage{array}                      
\usepackage{soul}                       
\usepackage[T1]{fontenc}                
\usepackage[hidelinks]{hyperref}        

\usepackage[separate-uncertainty=true,multi-part-units=single]{siunitx}
\DeclareSIUnit\sq{\ensuremath{\Box}}
\DeclareSIUnit\bar{bar}
\DeclareSIUnit\angstrom{\text{Å}}
\DeclareSIUnit{\atpercent}{at\%}

\newcolumntype{P}[1]{>{\centering\arraybackslash}p{#1}}
\hypersetup{
    colorlinks,
    linkcolor={red!50!black},
    citecolor={blue!50!black},
    urlcolor={blue!80!black}
}

\newcolumntype{L}[1]{>{\raggedright\arraybackslash}p{#1}}
\newcolumntype{C}[1]{>{\centering\arraybackslash}p{#1}}
\newcolumntype{R}[1]{>{\raggedleft\arraybackslash}p{#1}}

\begin{document}
\title{The effect of niobium thin film structure on losses in superconducting circuits}

\author{Maxwell Drimmer}
\thanks{These authors contributed equally to the work}
\affiliation{Department of Physics, ETH Z\"{u}rich, 8093 Z\"{u}rich, Switzerland}
\affiliation{Quantum Center, ETH Z\"{u}rich, 8093 Z\"{u}rich, Switzerland}
\author{Sjoerd Telkamp}
\thanks{These authors contributed equally to the work}
\affiliation{Department of Physics, ETH Z\"{u}rich, 8093 Z\"{u}rich, Switzerland}
\affiliation{Quantum Center, ETH Z\"{u}rich, 8093 Z\"{u}rich, Switzerland}
\author{Felix L. Fischer}
\thanks{These authors contributed equally to the work}
\affiliation{Department of Physics, ETH Z\"{u}rich, 8093 Z\"{u}rich, Switzerland}
\affiliation{Quantum Center, ETH Z\"{u}rich, 8093 Z\"{u}rich, Switzerland}
\author{Ines C. Rodrigues}
\affiliation{Department of Physics, ETH Z\"{u}rich, 8093 Z\"{u}rich, Switzerland}
\affiliation{Quantum Center, ETH Z\"{u}rich, 8093 Z\"{u}rich, Switzerland}
\author{Clemens Todt}%
\affiliation{Department of Physics, ETH Z\"{u}rich, 8093 Z\"{u}rich, Switzerland}
\affiliation{Quantum Center, ETH Z\"{u}rich, 8093 Z\"{u}rich, Switzerland}
\author{Filip Krizek}
\affiliation{Institute of Physics, Czech Academy of Sciences, Cukrovarnick\'a 10, 162 00 Praha 6, Czech Republic}
\author{Dominik Kriegner}
\affiliation{Institute of Physics, Czech Academy of Sciences, Cukrovarnick\'a 10, 162 00 Praha 6, Czech Republic}
\author{Christoph M\"uller}
\affiliation{Institute of Physics, Czech Academy of Sciences, Cukrovarnick\'a 10, 162 00 Praha 6, Czech Republic}
\author{Werner Wegscheider}%
\affiliation{Department of Physics, ETH Z\"{u}rich, 8093 Z\"{u}rich, Switzerland}
\affiliation{Quantum Center, ETH Z\"{u}rich, 8093 Z\"{u}rich, Switzerland}
\author{Yiwen Chu}%
\affiliation{Department of Physics, ETH Z\"{u}rich, 8093 Z\"{u}rich, Switzerland}
\affiliation{Quantum Center, ETH Z\"{u}rich, 8093 Z\"{u}rich, Switzerland}

\date{\today}

\begin{abstract}
\noindent The performance of superconducting microwave circuits is strongly influenced by the material properties of the superconducting film and substrate. While progress has been made in understanding the importance of surface preparation and the effect of surface oxides, the complex effect of superconductor film structure on microwave losses is not yet fully understood. In this study, we investigate the microwave properties of niobium resonators with different crystalline properties and related surface topographies. We analyze a series of magnetron sputtered films in which the Nb crystal orientation and surface topography are changed by varying the substrate temperatures between room temperature and 975 K. The lowest-loss resonators that we measure have quality factors of over $10^6$ at single-photon powers, among the best ever recorded using the Nb on sapphire platform. We observe the highest quality factors in films grown at an intermediate temperature regime of the growth series (550 K) where the films display both preferential ordering of the crystal domains and low surface roughness. Furthermore, we analyze the temperature-dependent behavior of our resonators to learn about how the quasiparticle density in the Nb film is affected by the niobium crystal structure and the presence of grain boundaries. Our results stress the connection between the crystal structure of superconducting films and the loss mechanisms suffered by the resonators and demonstrate that even a moderate change in temperature during thin film deposition can significantly affect the resulting quality factors. 

\end{abstract}
\maketitle

\section{Introduction}
The effort to build large-scale quantum processors has emphasized how studying materials can improve the lifetime and coherence of quantum systems~\cite{Murray2021,Siddiqi2021,deLeon2021}. Superconducting microwave resonators are often used to investigate loss in superconducting qubits because they are generally simpler to fabricate and measure while all sources of relaxation and decoherence affecting the performance of a resonator will also impact the performance of a qubit made from the same material~\cite{McRae2020}. Additionally, there are other applications that benefit from superconducting resonators with low levels of microwave dissipation, such as parametric amplifiers~\cite{aumentado2020superconducting}, quantum sensors~\cite{de2013near}, and microwave kinetic inductance detectors (MKIDs)~\cite{day2003broadband} used for astronomy and particle physics~\cite{ulbricht2021applications}. 

Measurements of superconducting resonators can distinguish between different sources of loss by determining the losses dependence on temperature and applied power. Dissipation at low microwave power and low temperature is especially relevant for quantum information processing~\cite{kjaergaard2020superconducting}. In this regime, a common limiting loss mechanism is dielectric loss due to two-level systems (TLS), and extensive work has been conducted in different material systems to locate and quantify this loss~\cite{Müller2019}. Oxides and impurities in close vicinity to the superconductor are a commonly observed cause of degradation of superconducting properties as they can host TLS~\cite{Altoe2022,Verjauw2021}. Many investigations attempting to reduce TLS loss have focused on the removal of unwanted oxides and impurities from different interfaces through etching or cleaning steps. Pre-deposition cleaning steps of the substrate to avoid metal-substrate interface losses~\cite{Wisbey2010, Sandberg2012, Megrant2012, Quintana2014, Bruno2015, Earnest2018}, as well as post-deposition processing steps of both substrate and deposited metal to reduce losses at the exposed surfaces~\cite{Goetz2016, Burnett2016, Lock2019, Verjauw2021, Kowsari2021, Altoe2022, Alghadeer2023, Crowley2023} have been investigated intensively. However, the connection between the superconductor film structure, which is determined by the deposition process, and the microwave loss suffered by circuits is not as well understood. Differences in the deposition process have been previously studied by some authors~\cite{Megrant2012,Dominjon2019,place2021new} who also report changes in microwave properties, but an accurate description of the effect of film crystal properties on TLS loss as well as other loss mechanisms has not yet been established.

Our study focuses on niobium, commonly used in the superconducting circuits community~\cite{Murray2021} because it has the highest critical temperature and critical magnetic field of any elemental superconductor~\cite{kittel2005introduction}. Furthermore, mono-crystalline Nb growth is possible at sufficiently high temperature and optimised growth conditions~\cite{Dominjon2019}. Sapphire substrates offer good thermal as well as chemical stability and have a small lattice mismatch with the Nb lattice, which allows for epitaxial growth~\cite{Wildes2001}. These key properties make Nb on sapphire suitable material platform to study the effect of crystallinity changes on superconducting resonator performance. 

We present a study in which we systematically investigate the effect of the Nb crystal properties and interface epitaxy on the behavior and performance of superconducting microwave resonators. In the first part of this study, we investigate the influence of the Nb growth temperature on the film crystallinity as well as surface morphology and interface epitaxy towards the sapphire substrate. The differences that we observe are then linked to the DC superconducting transport characteristics. In the second part, we characterize coplanar waveguide (CPW) resonators and explore their low-temperature microwave properties. Specifically, we investigate the relationship between thin film crystallinity as well as general morphology and the single-photon quality factors that we measure. Furthermore, we study power-independent losses and relate them to the grain structure of the superconducting films.

\section{Experimental design}

For this study, five thin film depositions were performed at various temperatures while all of the other deposition conditions were kept the same. The Nb thin films were sputtered in an Ultra High Vacuum (UHV) deposition chamber using DC magnetron sputtering~\cite{Todt2023}. The base pressure of the sputtering chamber was below $2 \times 10^{-10}$~mBar, indicating a low impurity background. The pressure during the deposition was 8.8$\times10^{-3}$~mBar and the distance between the 2 inch Nb target and the substrate was 110 mm. The power applied was 125 W corresponding to a stable growth rate of 0.27 nm/s at room temperature determined from in-situ Quartz Crystal Microbalance measurements. The effective growth rate decreased linearly with increasing temperature due to partial re-evaporation of adatoms from the surface. After growth, the thickness was recalibrated using profilometry and Transmission Electron Microscopy (TEM) measurements and the growth time was adjusted to achieve a thickness of around 185 nm for all films. The substrate temperature shown in~\autoref{tab:2} was set based on pyrometer readings of the temperature of the middle of the wafer. 

The Nb films were grown on epi-ready 420 um thick, (0001) sapphire wafers with a diameter of 2 inches. The substrates were pre-baked at 500 K for 8 hours in the load lock and subsequently outgassed for 20 minutes at growth temperature in the UHV chamber before growth. 

An overview of the growth parameters for all wafers is shown in~\autoref{tab:2}. 

\begin{table}[H]
    \centering
    \begin{tabular}{c|c|c|c|c|c}
          Wafer & \textbf{A}&  \textbf{B} & \textbf{C} & \textbf{D} &  \textbf{E}\\
         \hline
          Temperature [K] & 300 & 550 & 745 & 900 & 975 \\
         Pressure [mBar$\times10^{-3}$]  & 8.8 &8.8& 8.8& 8.8& 8.8\\      
         Power [W] & 125 & 125&  125 & 125&125\\

    \end{tabular}
    \caption{Overview of the sputtering conditions of the five wafers investigated in this study.}
    \label{tab:2}
\end{table}

From this series of Nb thin films, we fabricated chips containing CPW resonators~\cite{Pozar2011}. One chip features four open-circuit $\lambda / 2$-resonators of different lengths which are coupled to a central waveguide in the "hanger" or "notch" geometry \cite{McRae2020, Probst2015} (see Appendix~\autoref{geom_design}). We pattern this design using a photo-lithography process followed by a $\text{SF}_6$-based reactive-ion etch. After etching, the excess resist is stripped and the wafer is diced into chips which are packaged and measured in a dilution refrigerator. For more details about fabrication and measurement, see Appendices~\ref{sec:ResonatorFab} and~\ref{sec:MeasurementDetails}. 

\section{Results}
\subsection{Nb thin film crystal properties}
\label{material}
The crystal properties of a sputtered superconducting film can be affected by varying the substrate temperature during the deposition. If the substrate temperature is relatively low, deposited material forms a poly-crystalline film consisting of small crystal grains without a preferred orientation. At higher temperature, adatoms on the surface are supplied with increased energy that results in enhanced surface mobility. This leads to self-assembly into larger grains that are oriented in the most energetically favorable way, which in turn has an effect on the film morphology as described by a Thornton Zone Diagram~\cite{Thornton1988,Flynn1988}.

\begin{figure}
 
    \centering
    \includegraphics[width = 0.95\linewidth]{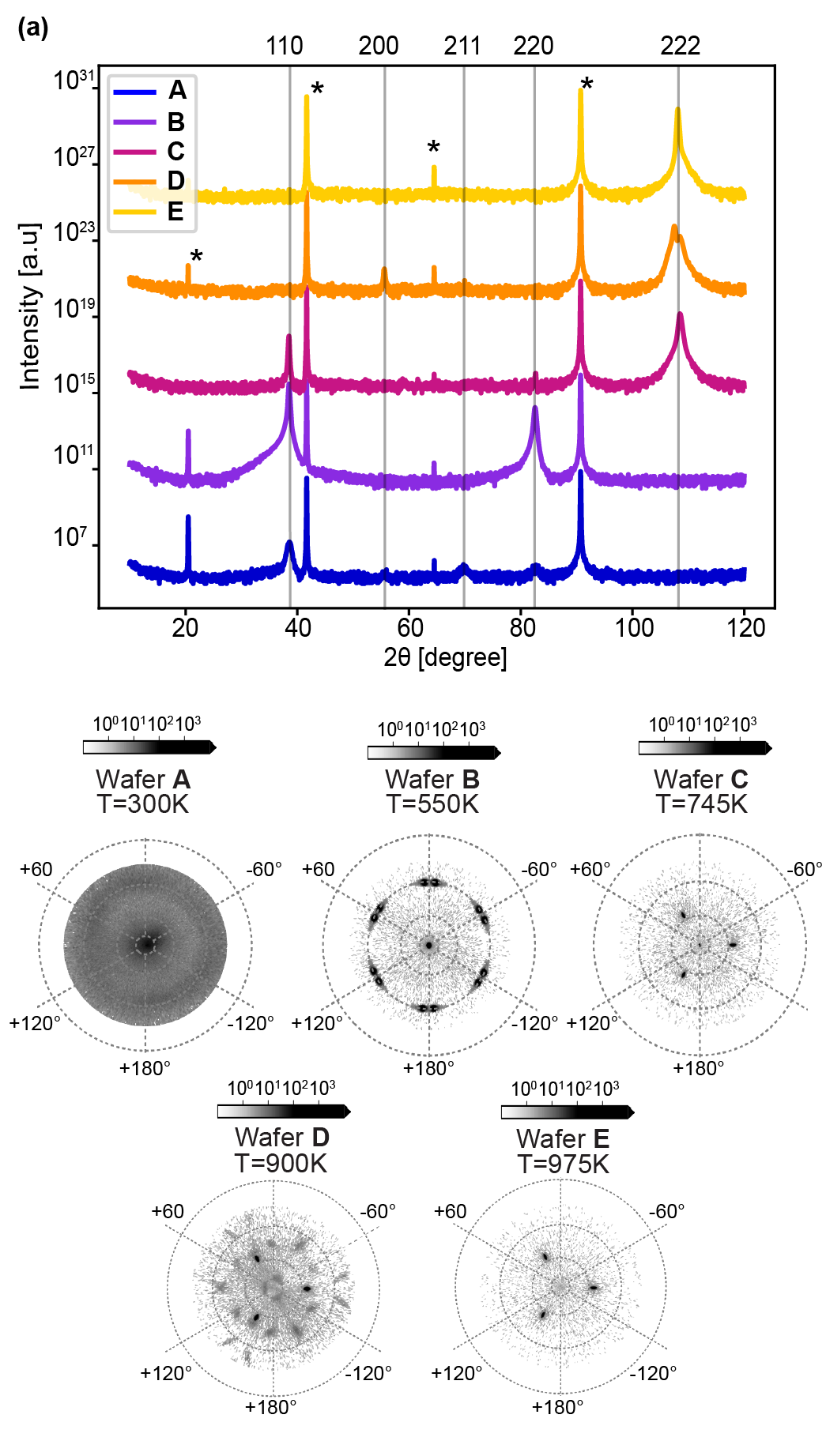}
    \caption{(a) XRD radial scans of the wafers listed in Table~\ref{tab:2}. The traces are offset with respect to each other for clarity. The sharp peaks that are associated with the mono-crystalline sapphire substrate are labeled with a star.  (b) XRD polefigures of the 110 Bragg peak showing the orientation distribution of the [110] direction for films deposited at various temperatures.}
    \label{XRD1}
\end{figure}

\autoref{XRD1}~(a) shows X-Ray diffraction (XRD) traces of the Nb films listed in~\autoref{tab:2}. Various Nb diffraction lines are indicated in the figure. Film \textbf{A} that was deposited at room temperature shows signatures of all the commonly observed diffraction lines of poly-crystalline Nb, namely 110, 200, 211, and 220~\cite{Edwards1951}. 

We observe a clear crossover in crystallinity from poly-crystalline at room temperature (Wafer \textbf{A}) to mono-crystalline at 975 K (Wafer \textbf{E}). The crystallographic changes become apparent at 550 K (Wafer \textbf{B}), where the [211] crystal orientation disappears and exclusively the [110] out-of-plane orientation is observed. At 745 K (Wafer \textbf{C}) the dominant out-of-plane crystal orientation changes to the [111] direction, which is not visible in Wafers \textbf{A} or \textbf{B}. Finally, at 975 K the 222 peak is the only visible peak associated with Nb in the XRD spectrum which indicates that we have grown a mono-crystalline film with [111] out-of-plane orientation. We do not directly observe the 111 peak because it is a forbidden reflection for Nb.

This crystallographic crossover from poly-crystalline to mono-crystalline is confirmed by the XRD polefigures shown in~\autoref{XRD1}~(b). In this figure, a stereographic projection of the diffraction signal from the {110} lattice planes is shown, from which we can also asses the in-plane crystal orientations. The signal in the center of the figure corresponds to an out-of-plane [110] orientation. For Film \textbf{A}, the figure shows an almost homogeneous distribution of diffraction intensity. This indicates that there is no directional preference of the crystal grains and is consistent with a poly-crystalline film. Starting at 550 K, dominant crystal grain orientations are formed as indicated by discrete features in the polefigure. At 745 K, highly intense diffraction spots appear which are consistent with the [111] out of plane orientation. The three observed peaks correspond to the three-fold rotation axis around the [111] direction. However, a weak contribution in the center of the polefigure remains at 745 K as well as some minor contributions at 900 K that are not aligned with the dominant orientation. At 975 K, a completely mono-crystalline structure with [111] out-of-plane orientation is formed and no other orientations can be detected within the probed area of several mm$^2$.

\begin{figure}
    \centering
    \includegraphics[width = 0.95\linewidth]{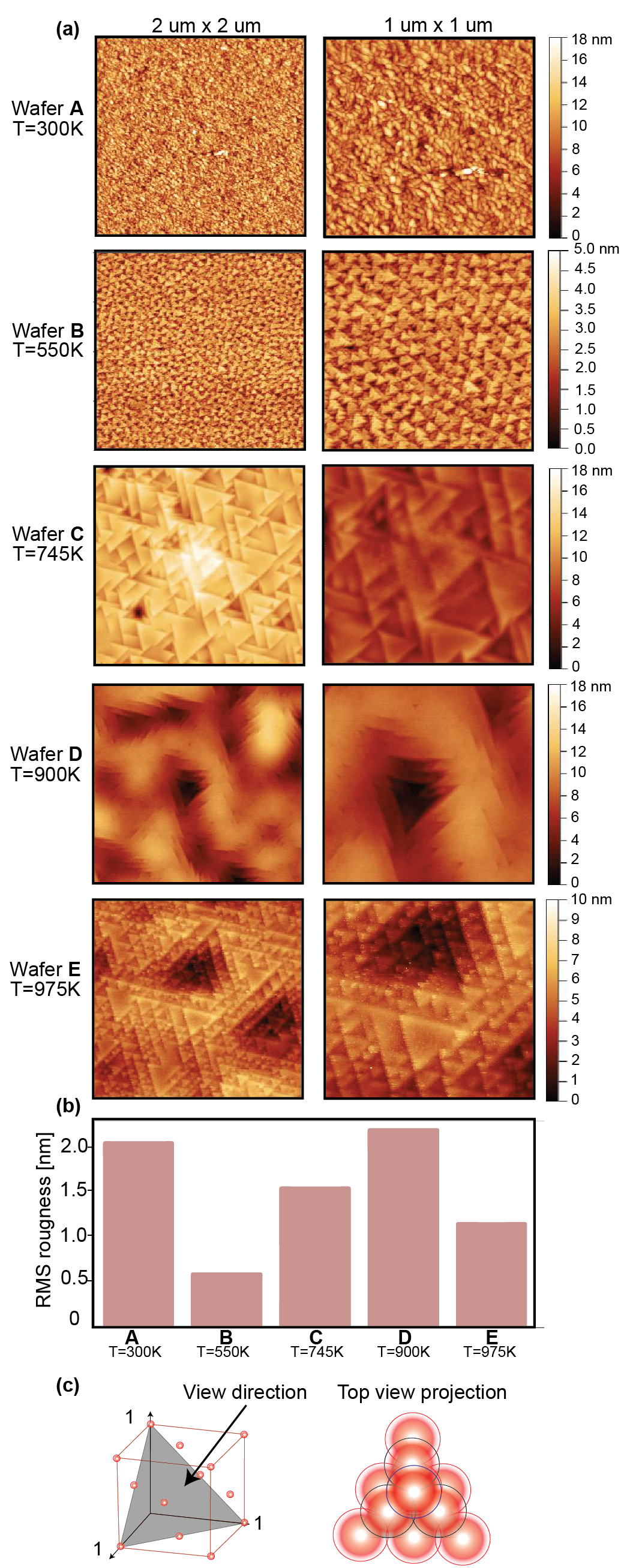}
    \caption{An overview of the effect of deposition temperature on Nb surface morphology studied by AFM topography images. In (a), Wafer \textbf{A} shows the commonly observed randomly oriented grain structure of Nb while the wafers grown at higher temperature start to display a triangular pattern that can be associated with the [111] direction viewed from the top (as indicated in (\textbf{c})). Figure (b) shows the $R_q$ surface roughness parameter associated with each wafer.}
    \label{AFM}
\end{figure}

The deposition temperature also influences the surface roughness of the sputtered Nb film. The surface topography was measured using atomic force microscopy (AFM) for each individual wafer directly after growth and is shown in~\autoref{AFM}~(a). The surface of the wafer grown at room temperature clearly consists of small grains with no apparent preferential orientation. This grain structure of Nb has been frequently observed on a large variety of substrates such as Si~\cite{Imamura1992} and GaAs~\cite{Todt2023}. A distinct triangular pattern appears at 550 K that becomes more and more evident with increasing temperature. The schematic in~\autoref{AFM}~(c) illustrates how a top view of the Nb (111) lattice planes could result in the observed triangular geometry. Our data suggests that the crystal domains grow in size and consequently change the surface roughness and morphology in a complex process of self-assembly during growth. The effect on the root-mean-square (RMS) surface roughness calculated from the 2$\times$2~$\mu$m AFM areas is plotted in~\autoref{AFM}~(b). It is evident from this plot that Wafer \textbf{B} has the lowest surface roughness.

\begin{figure*}
 
    \centering
    \includegraphics[width = 1\linewidth]{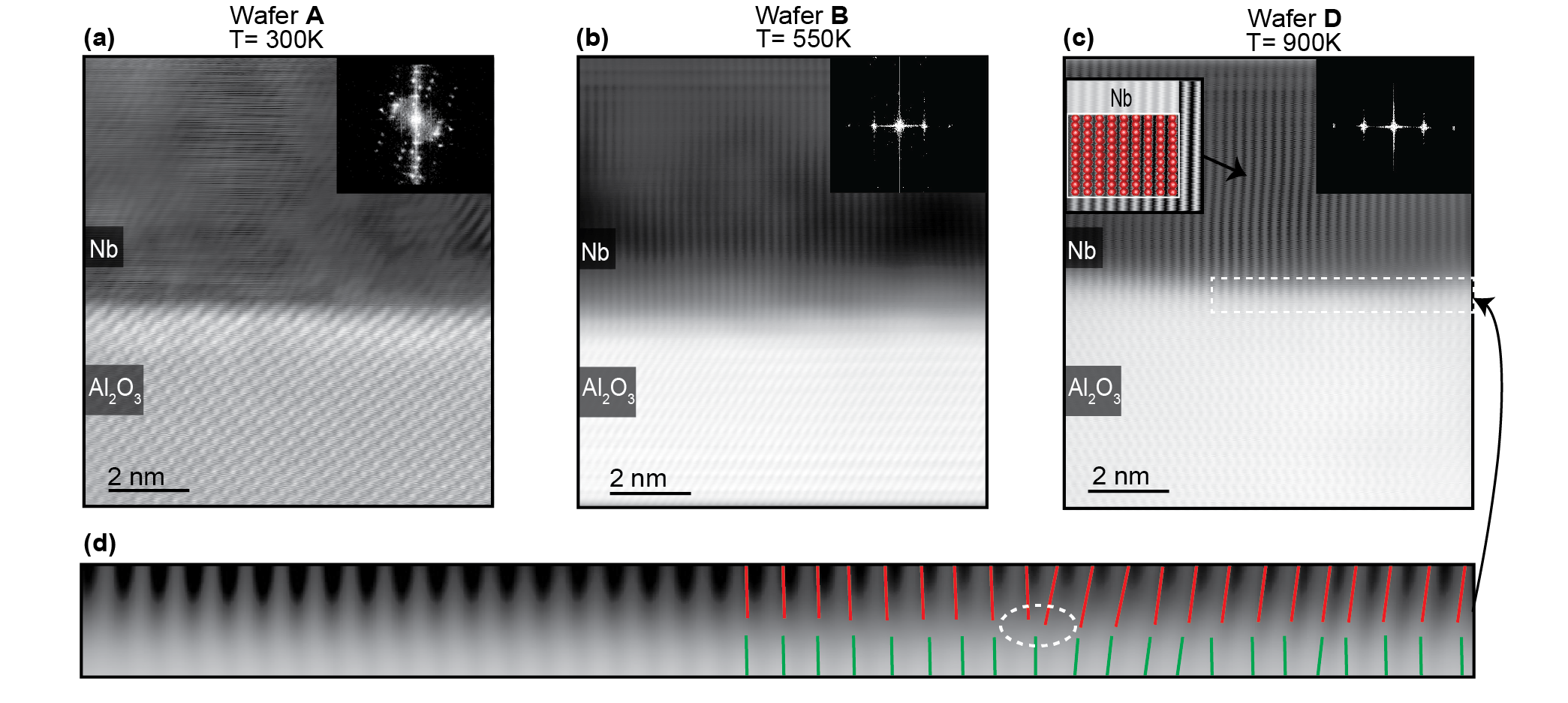}
        \caption{STEM images of Nb to sapphire interfaces of wafers grown at various temperatures. (a) A bright field STEM image in which a high-quality non-epitaxial interface is observed between the sapphire substrate and the poly-crystalline Nb film grown at room temperature (Wafer \textbf{A}). (b) A similar image of the interface between the Nb grown at 550 K and the sapphire substrate (Wafer \textbf{B}). In (c) the interface for Wafer \textbf{D} including a crystallographic model showing the Nb crystal lattice is shown. The zone axis differs between the images. In (d), a Bragg filtered image of a portion of the interface in (b) is shown in which the formation of sporadic misfit dislocations is indicated.} 
    \label{interface}
\end{figure*}

Another key feature that is known to have an effect on superconducting microwave resonator performance is the quality of the interface between the superconductor and the substrate~\cite{Wang2015}. We investigated the Al$_{2}$O$_{3}$-Nb interface of Wafers \textbf{A}, \textbf{B} and \textbf{D} using Scanning Transmission Electron Microscopy (STEM). In~\autoref{interface}~(a), it can be observed that for films grown at room temperature, a high-quality non-epitaxial interface is obtained that does not show any signs of oxides or chemical intermixing. The poly-crystalline character of the Nb layer is further substantiated by the ring-shaped pattern observed in the fast Fourier transform (FFT) image in the top right corner. At intermediate growth temperatures, it can be seen in~\autoref{interface}~(b) that an interface with a high degree of epitaxy between the sapphire (0001) and the Nb (110) crystals forms. Some variation in grain alignment was observed on the length scale of the lamella window ((1.5 $\mu$m) for Wafer \textbf{B}. In~\autoref{interface}~(c), the Nb to sapphire interface of Wafer \textbf{D} is shown. For this wafer, no grain boundaries were observed on the length scale of the lamella window and the Nb (111) to sapphire interface seems fully epitaxial. Therefore, we conclude that the (200) peak observed in the XRD measurements shown in~\autoref{XRD1} is only a very minor portion of the total crystal structure. A crystallographic model of the Nb lattice made using crystal visualization software (VESTA~\cite{VESTA}) is overlaid with a zoomed-in part of the figure and shows agreement with the image. The intense highly symmetrical features of the FFT in the in-set of~\autoref{interface}~(c) clearly confirm the single crystalline nature and high degree of epitaxy in the measured portion of the film. The strain relaxation by the formation of a periodic array of misfit dislocation can be deduced from the Bragg filtered image of Wafer \textbf{D} shown in~\autoref{interface}~(c). 
\subsection{DC transport characterization}

Before fabricating the resonators, the critical temperature and magnetic field of the superconducting Nb films were measured to benchmark the quality of the growth. Low-temperature DC transport measurements were performed in a Van der Pauw geometry using standard lock-in techniques to obtain the T$_c$ and B$_c$ of our wafers. All of the films have critical temperatures around $9.25\pm0.1$K, which is close to the maximum bulk value of Niobium of 9.3 K~\cite{Roberts1976}. We found that the critical magnetic field was consistently higher for the films with a lower degree of crystallinity. This can be attributed to superconducting vortices that are formed in the mixed state ($B$>$B_{c1}$) and are pinned on grain boundaries in the material \cite{Dobrovolskiy2012}. More details on these measurements can be found in Appendix~\ref{sec:dcTransport}. 

The mean free path $l_\mathrm{MFP}$ and the Ginzburg-Landau (GL) coherence length $\xi_{GL}(0)$ are determined from these measurements since they can be used to identify possible sources of decoherence. Using the phase diagram shown in supplementary~\autoref{BcTc}, we extrapolate the GL B$_{c2}$ value to find the critical magnetic field at 0 K. The GL coherence length was calculated using~\autoref{Glequation} and the mean free path is determined by measuring the wafer's resistivity at 10 K and using~\autoref{mfp}. The resulting values are shown in~\autoref{tab:1}. 

\begin{table}[H]
    \centering
    \begin{tabular}{c|c|c|c|c|c}
          & Wafer \textbf{A}& Wafer \textbf{B} & Wafer \textbf{C} & Wafer \textbf{D} & Wafer \textbf{E}\\
         \hline
          $\rho_\mathrm{10K}$ [$\mu\Omega$cm] & 4 & 2.3 & 1.1 & 0.77 &0.74\\
          $l_\mathrm{MFP}$ [nm] & 9 & 16 & 34& 48&50\\
                  $\xi_\mathrm{GL}(0)$ [nm] & 14 &20 & 23 & 23 & 23 \\
         
    \end{tabular}
    \caption{The residual resistivity at 10 K $\rho_\mathrm{10K}$, the mean free path length $l_\mathrm{MFP}$ derived from $\rho_\mathrm{10K}$, and the Ginzburg-Landau coherence length $\xi_\mathrm{GL}(0)$ determined from the critical magnetic field values shown in~\autoref{BcTc}.}
    \label{tab:1}
\end{table}

\subsection{Microwave characterization}

The losses of microwave resonators were measured by analyzing the S$_{21}$ scattering response and extracting the linewidth and resonance frequency of each resonance. The different physical mechanisms which contribute to the observed loss rate can be distinguished by varying the applied microwave power and temperature~\cite{McRae2020}. According to similar studies in literature, we expect dielectric losses due to TLS to be the dominant source of loss in the low temperature and single-photon power regime~\cite{Verjauw2021, Altoe2022}. In addition to losses from TLS, which scale strongly with applied microwave power, we consider loss mechanisms that do not depend on power. These include temperature-dependent losses due to thermal quasiparticles, but also temperature-independent losses, e.g. due to external pair-breaking radiation or parasitic modes within the resonators~\cite{Bruno2015,Barends2010,Verjauw2021,Zheng2022}. Therefore, we include $\delta_\mathrm{PI}(T)$, a power-independent (PI) loss term, in our model:
\begin{equation}
\label{eq:tls_fit}
    \frac{1}{Q_\mathrm{int.fit}(T,P)} = \frac{1}{Q_\mathrm{int}^\mathrm{TLS.full}(T,P)} + \delta_\mathrm{PI}(T)
\end{equation}
Other authors have employed this practice in different variants, some explicitly modeling the temperature dependence of PI losses~\cite{Crowley2023,Goetz2016}, some splitting the total loss into a high-power and a low-power regime~\cite{Earnest2018,Calusine2018,Altoe2022,Gao2022}. The full expression for our TLS loss model $\frac{1}{Q_{\mathrm{int}}^{\mathrm{TLS.full}}(T,P)}$ is presented in Appendix~\ref{eq:TLS_full}.

\begin{figure}
    \centering
    \includegraphics[width = \linewidth]{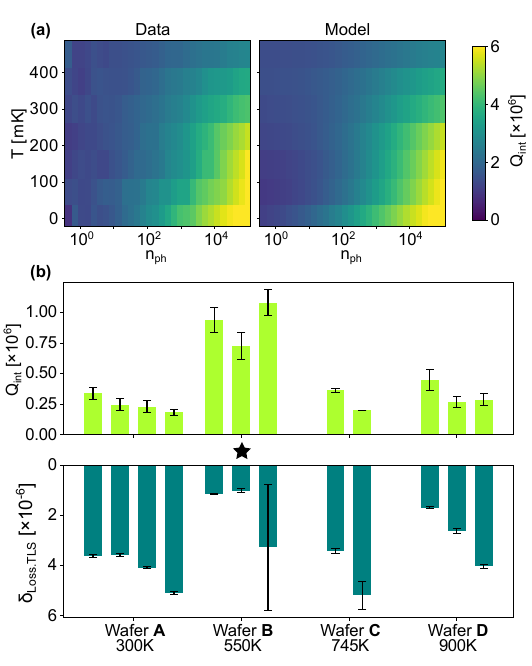}
    \caption{(a) Temperature and power dependence measured of the quality factor of the starred resonator from Wafer \textbf{B} marked in (b). Data is shown on the left while the model that was fit to that data is on the right. (b) The top panel is the low-temperature, single-photon power internal quality factors extracted from all resonators that we measured, excluding those with fabrication defects. The bottom panel shows the TLS losses extracted from the power and temperature dependence of the internal quality factor for each of the resonators. The TLS loss determined from fitting one of the resonators from Wafer \textbf{B} was subject to large error, because at ten photons in the resonator the internal quality factor exited the critical coupling regime (see Appendix~\ref{Apdx:Coupling_regimes}).}
    \label{TLS_figure}
\end{figure}

\autoref{TLS_figure}~(a) compares a measurement of the internal quality factor of one resonator from Wafer \textbf{B} to \autoref{eq:tls_fit}. The excellent agreement indicates that our model captures the temperature and power dependence of the measured quality factors (see  Appendix~\ref{apdx:fittingProcedures} for the same comparison for all resonators shown in~\autoref{TLS_figure} (b) ). As predicted by TLS theory, the internal quality factor is highest at high power and has a non-straightforward temperature dependence originating from the competition of TLS saturation and TLS interactions~\cite{Goetz2016,Müller2019,Crowley2023}. Additionally, fitting to our model allows us to distinguish TLS loss from PI loss. In~\autoref{TLS_figure} (b) we show the low-temperature single-photon quality factors for each resonator that we measured together with the extracted TLS losses. We excluded any resonators with visible fabrication defects. The yield of working resonators varied for each chip but was always $\geq 50\%$. Resonators from Wafer \textbf{B}, grown at 550 K, show substantially higher quality factors and lower TLS losses than resonators grown at the other temperatures. The highest single-photon $Q_\mathrm{int}$ we measured was $1.1\pm0.1\times10^6$, which to our knowledge is among the highest ever measured single-photon $Q_\mathrm{int}$ for a niobium on sapphire resonator~\cite{Geerlings2012,wang2024}. 

We also fabricated resonators from Wafer \textbf{E}, but observed more than one order of magnitude lower internal quality factors than for the other wafers. We attribute this to the high number of defects that exist in wafers grown at 975 K (more detail can be found in Appendix~\ref{apdx:defects}). The formation of these defects is likely a result of surface adhesion problems during film deposition at these high temperatures. We observe only small variations in internal quality factor between Wafers \textbf{A}, \textbf{C}, and \textbf{D} that are on the order of variations between resonators from the same wafer. 

\begin{figure}
    \centering
    \includegraphics[width = \linewidth]{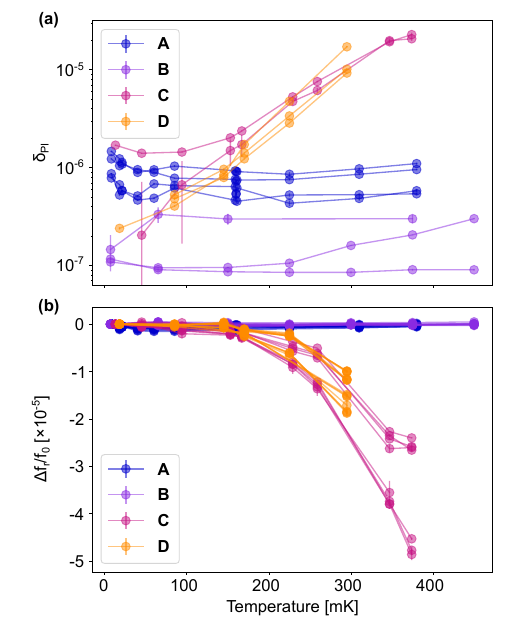}
    \caption{Power-independent loss (a) and fractional frequency shift (b) of each resonator measured in~\autoref{TLS_figure}~(b) as a function of temperature.}
    \label{PI_fr_figure}
\end{figure}

The fitted PI losses are shown as a function of temperature in~\autoref{PI_fr_figure} (a). We observe a clear temperature dependence of the PI loss for Wafers \textbf{C} and \textbf{D} while the PI loss appears constant for Wafers \textbf{A} and \textbf{B}. We further observe differences in the temperature-independent contribution to $\delta_\mathrm{PI}$ that we attribute to slight differences in the measurement environment. These differences are rather small compared to the temperature-dependent variations and compared to the loss from TLS. In \autoref{PI_fr_figure} (b), the temperature dependence of the fractional frequency shift $\Delta f_r/f_0 = (f_r(T)-f_r(T\approx0))/f_r(T\approx0)$ shows a similar trend, where $f_r$ is the resonator frequency extracted from fitting the raw data. This behavior will be further explored in the discussion section. The PI losses are smaller than the TLS losses for resonators from the Wafers \textbf{A} and \textbf{B} at all temperatures. For Wafers \textbf{C} and \textbf{D}, we observe that the PI loss dominates over the TLS loss at the highest temperatures that we measured.

\section{Discussion}\label{sec:discussion}

The differences that we observed between microwave measurements of our wafers can be interpreted within the context of material properties of our films presented in~\autoref{material}. From the XRD characterization shown in~\autoref{XRD1} it can be seen that the biggest change in crystallinity occurs between Wafers \textbf{A} and \textbf{B} and that comparatively little change occurs between Wafers \textbf{C} and \textbf{D}. This interpretation of the XRD data is in line with the values of $l_\mathrm{MFP}$ and $\xi_{GL}(0)$ shown in~\autoref{tab:1} which indicate that grain boundaries become less frequent in Wafers grown at larger temperatures. The AFM data displayed in~\autoref{AFM} demonstrates a less straightforward trend, but we can clearly see that Wafer \textbf{B} has a far lower surface roughness than the other films. This could be attributed to the observation that only the niobium film on Wafer \textbf{B} is primarily oriented in [110] direction while Wafers \textbf{C} and \textbf{D} are dominantly oriented in the [111] direction.

The power-dependant response of our resonators indicates that the dominant mechanism limiting our single-photon-power low-temperature internal quality factors is loss due to TLS. We achieve the lowest TLS loss for Wafer \textbf{B}, which features the lowest surface roughness, suggesting that the roughness of the film has an effect on TLS loss. The exact mechanism that describes how surface morphology affects the TLS loss in our study is unclear. In films with larger surface roughness, the TLS loss could be higher due to an increased volume of dielectric material or the presence of sharp surface features that concentrate the electrical field distribution at the surface. Furthermore, changes in surface morphology observed in the AFM or different dominant crystal orientations could favor the growth of one niobium oxide species over another. This would lead to a change in the TLS loss because certain oxide species are known to host higher densities of TLS~\cite{Murthy2022}.

It is however premature to correlate dielectric loss exclusively with the roughness of a film. For instance, changes in surface roughness between Wafers \textbf{C} and \textbf{D} are not clearly reflected in their dielectric losses. Furthermore, the average TLS loss of Wafer \textbf{A} is 1.5 times that of Wafer \textbf{D}, even though Wafer \textbf{D} shows a slightly higher surface roughness. To explain these differences, we suggest a separate contribution to the overall TLS loss that depends on grain structure. Oxides are known to penetrate metallic films along grain boundaries~\cite{halbritter2005transport}, potentially hosting additional TLS. Furthermore, disorder in the crystal due to the grain boundaries themselves could also lead to higher TLS loss~\cite{Müller2019}. In contrast, contributions from other common sources of TLS such as the substrate-air and metal-substrate interfaces are likely negligible in our study from qualitative arguments. The substrate-air interface is completely unchanged between different chips and the results of our STEM study convince us that the metal-substrate interface has no oxides and no visible structural damage for both high and low-temperature depositions.

Other authors who investigated the effect of different growth conditions on microwave dissipation report higher internal quality factors/T1 times for growths at 770 K~\cite{place2021new} or at 1070 K~\cite{Dominjon2019} compared to room-temperature growths. Furthermore, in other material systems, highly ordered films displayed lower losses than granular films~\cite{Megrant2012}. Both types of studies only compared the extremes of a continuous spectrum of growth conditions and grain structures. Our results do not contradict these findings, instead, we report a temperature "sweet spot" in our growth series where the combination of all the factors that affect TLS loss produces the highest low-energy quality factors.

We can also interpret our measurements to learn about sources of loss other than TLS. The temperature-dependant behavior shown in~\autoref{PI_fr_figure} is likely a result of thermally populated quasiparticles, since only this loss mechanism both increases the dissipation and decreases the resonance frequency of a resonator as a function of temperature~\cite{McRae2020}. In our low-temperature regime, the density of quasiparticles depends on the film's quasiparticle trapping rate which is related to the grain boundary density~\cite{Wang2014}. We argue that as the growth temperature exceeds 550 K, our wafers enter a regime where the quasiparticle trapping is reduced to an extent that we see an effect on resonator frequencies and losses. However, beyond Wafer \textbf{C} there does not seem to be a big difference in the trapping of quasiparticles. We speculate that the reason behind this behavior lies in changes in the trapping mechanism of thermal quasiparticles. The global amount of grain boundaries/scattering sites decreases from Wafer \textbf{A} to \textbf{D}, as indicated by the MFP increasing. The vortex trapping behavior resulting from these grain boundaries likely affects the films less as films become more mono-crystalline, as evidenced by the coherence length increasing from Wafer \textbf{A} to Wafer \textbf{C} and then staying constant through Wafer \textbf{E}. We theorize that Wafers \textbf{A} and \textbf{B} are relatively unaffected by thermal quasiparticles up to $\sim$500 mK because of quasiparticle trapping at grain boundaries. However, in Wafers \textbf{C} and \textbf{D} this trapping mechanism is no longer able to suppress thermal quasiparticles and as a result we observe power independent losses that increase with temperature.

\section{Conclusion}
In the first section of our study, we show that the crystallographic structure of Nb on sapphire films is drastically affected by changing the deposition temperature. This resulted in films ranging from a poly-crystalline structure (Sample \textbf{A}) at room temperature to a completely mono-crystalline character at the highest temperature of 975 K (Sample \textbf{E}). With increasing temperature, the films have fewer grain boundaries and the crystal domains undergo an increasing degree of ordering in the preferred [111] orientation. These differences in crystallinity also lead to variation in surface roughness, most notably in Wafer \textbf{B}, which has the lowest surface roughness. 

These differences in material properties lead to significant variations in the microwave behavior of our superconducting resonators. Specifically, we found the best single-photon quality factors of over one million in the resonators made from Wafer \textbf{B}, which was grown at an intermediate temperature of 550 K. These quality factors are state-of-the-art for Nb on sapphire resonators~\cite{Geerlings2012,wang2024}, and are within an order of magnitude of the state-of-the-art for niobium resonators on any substrate~\cite{Verjauw2021}. We believe that the TLS loss rate could be further reduced by optimizing the fabrication process by including steps like a HF etch to remove the processing oxide~\cite{Verjauw2021,Altoe2022,Crowley2023}, etching trenches~\cite{Calusine2018,Nersisyan2019,woods2019determining}, encapsulating the resonator~\cite{Alghadeer2023, bal2024systematic}, or optimizing the CPW geometry for low electric field participation at material interfaces~\cite{Calusine2018,woods2019determining,ganjam2023surpassing}. Our results could also be strengthened by measuring more resonators on more samples, as the TLS losses we report vary from resonator to resonator. 

This result indicates that TLS losses in Nb qubits on sapphire are reduced by increasing the substrate temperature by only 250 K above room temperature. A similar optimum might also exist for other material systems and are a subject for further investigation. Growing at moderate temperatures will significantly decrease the number of grain boundaries in the film while also leading to relatively low surface roughness. Our work indicates that these factors could reduce the TLS losses by up to an order of magnitude. A relatively small temperature increase during growth could be applied in many different sputtering systems, since it does not put as stringent requirements on the amount of background impurities in the system as the high-temperature growth of, for example, Wafer \textbf{D} does. 

We also observe a temperature dependence in our resonator measurements that we attribute to the degree of crystallinity of the films. Our data displays two distinct behavioral regimes which we suspect arise from the relation between the quasiparticle dynamics in the film and the grain boundary density. This could be an important consideration for studying quasiparticle trapping dynamics and could be further investigated by future experiments, for example those with a magnetic field degree of freedom. Finally, the experiment would additionally benefit from x-ray photoelectron spectroscopy (XPS) and TEM electron-energy-loss spectroscopy (EELS) analysis to determine the composition of the surface oxide for each wafer.

\section*{Acknowledgements}
We thank M. Bahrami Panah, I. Kladari\'c, H. Doeleman, T. Schatteburg, L. Michaud, E. Planz, S. Andresen, and Y. Yang for valuable discussions, ScopeM and specifically P. Zeng for lamella preparation by FIB and TEM support and M. Sousa (IBM) via the Binning and Rohrer Nanotechnology Center (BRNC) for TEM support. \textbf{Funding:} This project has received funding from the European Research Council (ERC) under the European Union’s Horizon 2020 research and innovation programme (Grant agreement No. 948047). This project was additionally supported by the Swiss National Science Foundation (SNSF), Czech Science Foundation (Grant No. 22-22000M), MEYS CR grant nr. CZ.02.01.01$\slash$00$\slash$22$\_$008$\slash$0004594, and Czech Academy of Sciences (project no. LQ100102201). \textbf{Author contribution:} M.D., S.T., and F.F. conceived the experiment. S.T. grew the wafers with technical support from C.T.. S.T. F.K., D.K., and C.M. performed and analyzed the XRD measurements. S.T. and F.K. preformed and analyzed the STEM measurements. S.T. and F.F. characterized the wafers using DC transport and AFM measurements. I.R. and F.F. designed the resonators. M.D. fabricated and packaged the devices. M.D. and I.R. constructed the microwave measurement setup. M.D. and F.F. performed the microwave measurements. F.F. performed the fitting and analyzed the data with support from M.D. and I.R.. M.D., S.T., F.F. interpreted the results and wrote the manuscript, which was revised by all authors. The work was supervised by Y.C. and W.W..


\appendix

\section{Growth Defects at High Temperatures}\label{apdx:defects}
At growth temperatures above 900 K, we observe high densities of point-like growth defects. We propose that reduced adhesion of Nb atoms at high growth temperatures is the cause of these defects. Any impurity or defect on the sapphire substrate surface can lead to localized spots where the Nb does not nucleate homogeneously and ultimately results in a hole in the film. The density of defects that we observed is always high at high growth temperatures but can vary from wafer to wafer. We observed this variation even though the substrate preparation was the same for all growths. While for wafers of type \textbf{D} grown at 900 K we can achieve low defect densities, this was not possible for wafers grown at 975 K (Wafer \textbf{E}). For this wafer, even after multiple attempts, the defect density remained high. Resonators that are fabricated on wafers with visibly high defect densities show orders of magnitude lower single-photon power low-temperature internal quality factors (1.0$\times10^4$ and 1.2$\times10^4$ for two resonators on Wafer \textbf{E}) compared to our other devices. We exclude these devices from our main study, as the mechanisms limiting them are exclusive for the highest temperatures and no clear trend is visible. 

\begin{figure}
    \centering
    \includegraphics[width = 1\linewidth]{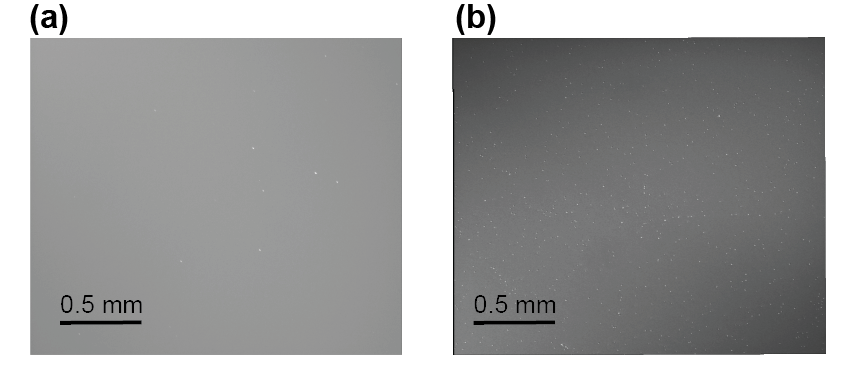}
    
    \caption{Optical microscope images of adhesion defects. 
     (a) shows a wafer with low defect density grown at room temperature. (b) shows a wafer with a very high defect density grown at 900 K.}
    \label{fig:defects_optical}
\end{figure}

A qualitative comparison of the density of these defects between a high-temperature and a low-temperature wafer is shown in~\autoref{fig:defects_optical}.

\section{Resonator Design}

\begin{figure}
    \centering
    \subfigure[]{
    \includegraphics[width = 1\linewidth]{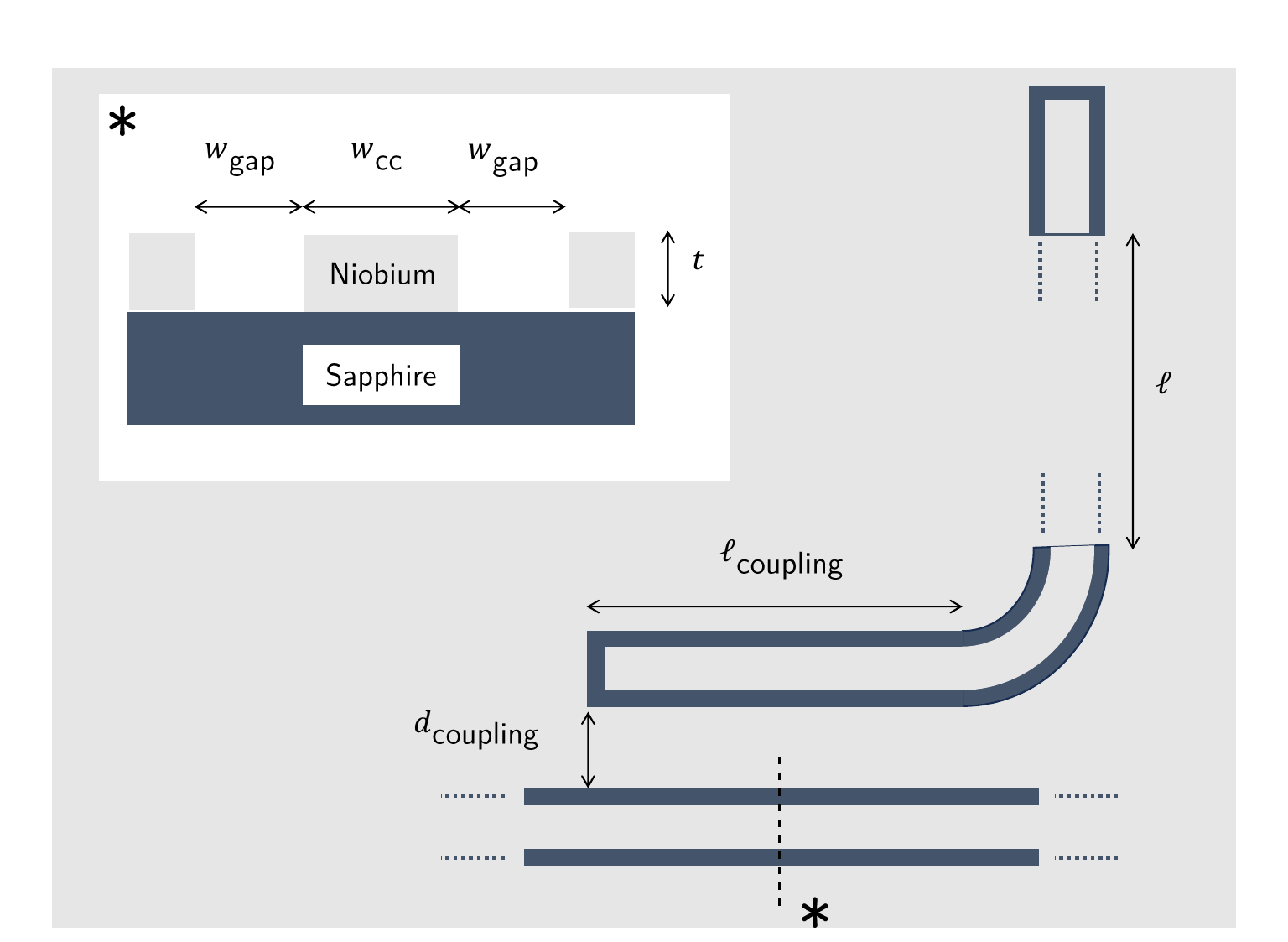}
    \label{geom_parameters}
    }
    \subfigure[]{
    \includegraphics[width = 0.9\linewidth]{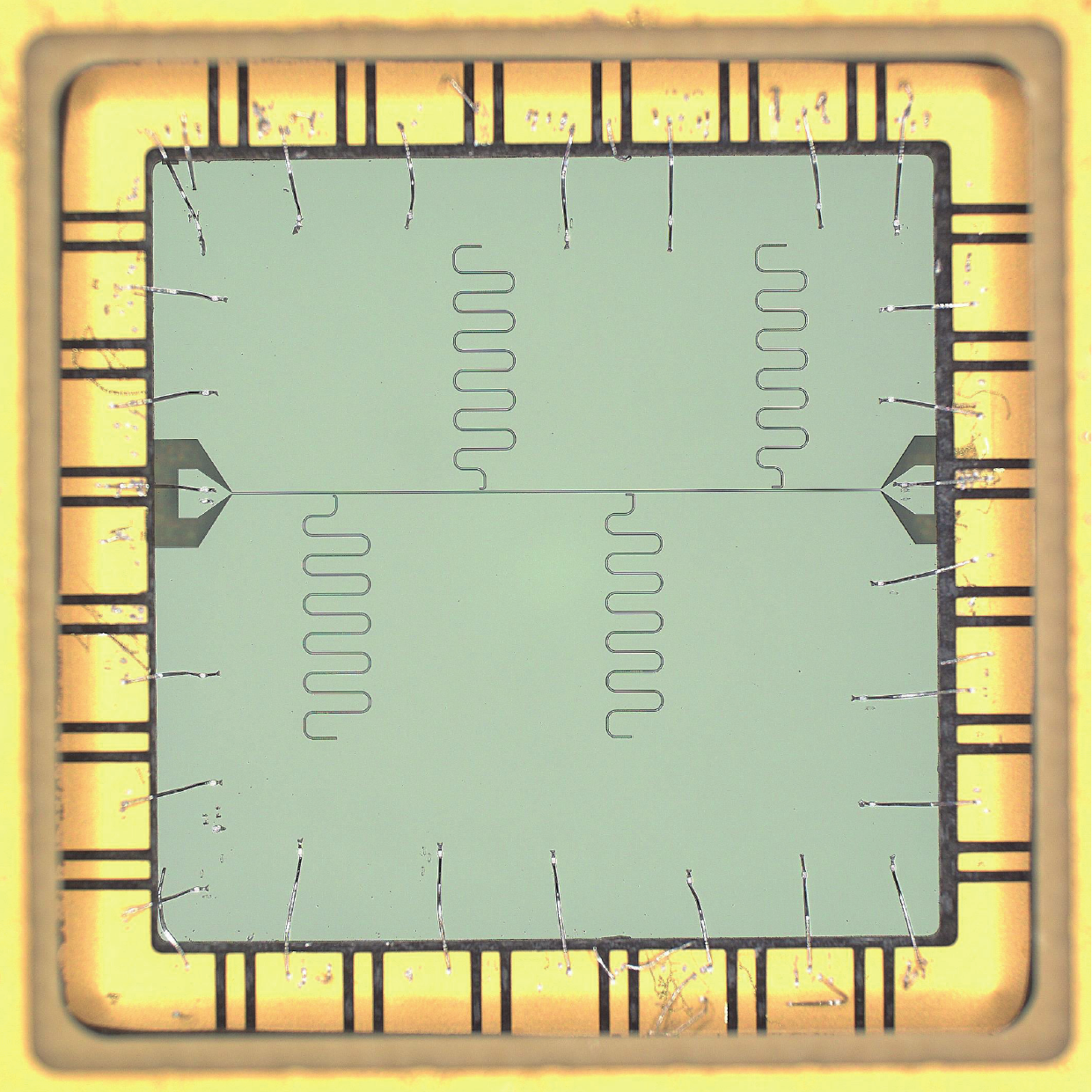}
    \label{chip_overview}
    }
    \caption{(a) A diagram with the important geometrical parameters in our chip design. (b) An optical microscope image of a chip from Wafer \textbf{B} after fabrication.}
    \label{geom_design}
\end{figure}

Superconducting microwave resonators serve as a proxy for the characterization of materials for application in superconducting qubits. A common implementation of a microwave resonator is a coplanar waveguide. A single resonator consists of a center conductor of width $w_\mathrm{cc}$ separated from a lateral ground plane by trenches of width $w_\mathrm{gap}$. Our chip design features four open-circuit $\lambda / 2$-resonators of lengths $l_i$ for $i = 1-4$ and one common feedline, contained in a $\SI{10}{mm} \times \SI{10}{mm}$ square chip. Each resonator is of a different length to separate their response in frequency space and to allow the measurement of all four resonators with one read-out signal. A length $l_\mathrm{coupling.i}$ of each resonator runs in parallel to the feedline with a spacing of $d_{coupling}$. In this layout, each resonator is coupled capacitively to the common feedline, with the coupling strength being controlled by varying $l_\mathrm{coupling.i}$ and $d_{coupling}$. A schematic of the design is shown in \autoref{geom_design}. For all chips, $w_{cc} = \SI{10}{\mu m }$, $w_{gap} = \SI{6}{\mu m}$, $d_\mathrm{coupling} = \SI{12}{\mu m}$, and $l_i = \left(8.782,8.160,7.609,7.120\right)$mm are kept constant between different chips. For this choice of $l_i$, we determined resonance frequencies of $\left(7,7.5,8,8.5\right)$GHz by simulating our design using AnsysEM. We vary the coupling length $l_\mathrm{coupling.i}$ in a range of $\SI{25}{\mu m} - \SI{350}{\mu m}$ between different chips and also within each chip. As a result, we measured coupling quality factors of $Q_\mathrm{c}\approx 4\times10^4 - 4\times10^6$.
Internal quality factors can only be determined with small error bars if they are within the critical coupling regime~\cite{Göppl2008} $Q_\mathrm{int}\sim Q_\mathrm{c}$ (see Appendix~\ref{Apdx:Coupling_regimes}). In our chip design, each resonator is coupled to the feedline with a different $l_\mathrm{coupling.i}$ such that the chip covers a broad range of possible internal quality factors. As a consequence, some resonators on our chips, for instance one resonator on Wafer \textbf{B} and Wafer \textbf{D}, were not in the critical coupling regime, which prevented us from extracting an accurate estimate of the internal quality factors.

\section{Resonator Fabrication}\label{sec:ResonatorFab}

The wafers are cleaned by being sonicated in acetone and then in isopropanol for two minutes each. Then the wafers are baked at 180$^{\circ}$ C for two minutes before we spin a $\sim$1.5 $\mu$m thick layer of AZ 5214E-JP image-reversal photoresist and bake at 105$^{\circ}$ C for one minute. We use an i-line soft contact masked photolithography process to expose the wafers and develop the resist in AZ 726 MIF for three minutes. 

\begin{figure}
    \centering
    \includegraphics[width = 1\linewidth]{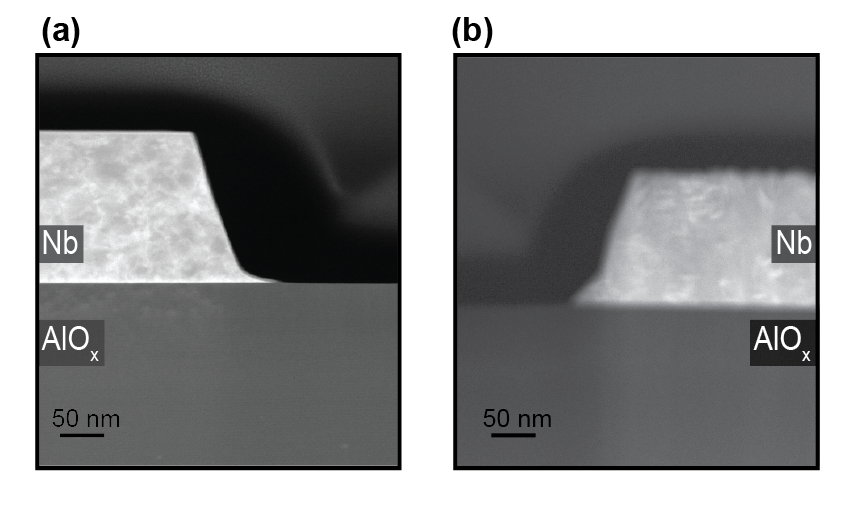}
    
    \caption{BF STEM images showing the cross-sectional profile of resonators after fabrication from Wafer \textbf{B} (a) and Wafer \textbf{A} (b).}
    \label{Etchprofile}
\end{figure}

After rinsing in water and drying, we load the wafers into an Oxford Instruments Plasmalab 80 Plus reactive-ion etcher. We flow five sccm of $\text{SF}_6$ and use an RF power of 100 W while illuminating the wafer with a helium-neon laser and monitoring the intensity of the reflected spot. When the intensity of the spot drops (typically after five minutes of etching), we continue etching the wafers for between 20-75 seconds in order to account for any anisotropy of the plasma. Subsequent STEM measurements of cross-sections (shown in~\autoref{Etchprofile}) of these resonators demonstrate the sidewall profile of this procedure. 

After etching, the excess resist is stripped in 70$^{\circ}$ C N-Methyl-2-pyrrolidone (NMP) for at least 30 minutes before the wafers were sonicated in NMP, then acetone, and finally isopropanol for one minute each and blown dry. A Disco DAD 3221 Dicing Saw is used to cut the wafer into 10x10 mm chips. We use small pieces of a 100 $\mu$m indium sheet to press our chip onto the corner pedestals of a QCage.24 sample holder from Quantum Machines and use aluminum wirebonds to connect the waveguide launchers to the transmission line of the package.

In the context of our conclusions regarding the effect of crystal structure on the microwave properties of our resonators, it is crucial to verify that our fabrication process was reproducible and did not lead to any unwanted variation in the measured resonators. Although effort was made to perform the fabrication process exactly the same for each chip, small variations in for example thickness of the films did require slightly different dry etching times. We show the cross-section of the etched trench for Wafers \textbf{A} and \textbf{B} in~\autoref{Etchprofile}. From this figure, it can be seen that a very similar etch profile is obtained in both chips. This result further substantiates that the fabrication process was indeed reproducible and the observed differences in microwave characterization are indeed related to the crystal structure rather than chip-to-chip variation in fabrication. 

\section{Microwave Measurement Details}\label{sec:MeasurementDetails}

We conduct microwave measurements in a dilution refrigerator. Our sample is thermalized to the base plate of the cryostat. The temperature of our sample is estimated from a thermometer on the base plate of the cryostat. The microwave response of the resonators is analyzed with a vector network analyzer (VNA) (\textit{Keysight}, P5004A). During the microwave characterization of our samples, we apply microwave signals with powers ranging from \SI{-80}{dBm} to \SI{10}{dBm} at frequencies between \SI{1}{GHz} and \SI{12}{GHz} and record the magnitude and phase of the transmitted signal. We repeat this procedure at temperatures between (\SI{8}{mK}, \SI{400}{mK}). Some measurements were repeated to check that the samples were well-thermalized and that the results were reproducible. 

\begin{figure}
 
    \centering
    \includegraphics[width = 0.7\linewidth]{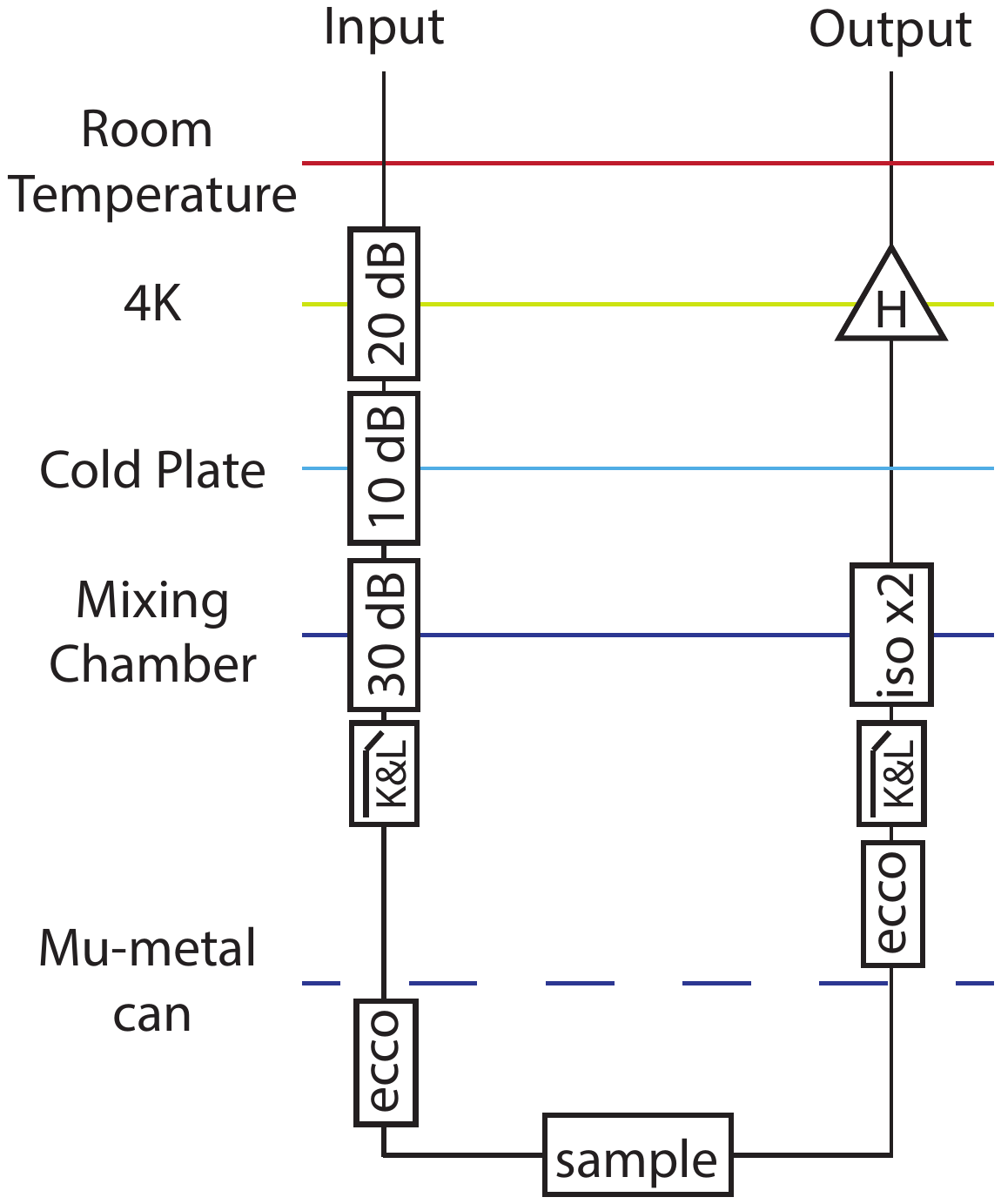}
    
    \caption{A schematic of the measurement chain inside of the dilution refrigerator. iso x2: Two isolators in series. K$\&$L: Low-pass filter. ecco: Eccosorb filter. H: HEMT.}
    \label{fig:measSetup}
\end{figure}

The input signal enters the fridge and is subject to 60 dB of attenuation installed on different stages of the cryostat as well as a low-pass filter (\textit{K$\&$L Microwave}, 6L250-12000/T26000-OP/O) and an eccosorb filter (\textit{Laird}, Eccosorb CR-110) before reaching the sample. The sample and eccosorb filter are inside of a mu-metal magnetic shield (\textit{Amuneal}, A4K Can). After the sample, the signal passes through another eccosorb and low-pass filter as well as two isolators (\textit{LNF}, ISISC4-12A) in series mounted at the mixing chamber plate. The output signal is further amplified at the 4 K stage by a high electron mobility transistor (HEMT) in addition to a room-temperature amplifier. A diagram of the setup is shown in~\autoref{fig:measSetup}.

In addition to the attenuation from the attenuators, the applied microwave power is further reduced by loss in the cables and attenuation due to microwave reflection at the connectors. To find the total attenuation of the input line we perform a power calibration measurement~\cite{Rodrigues2021}. While measuring over a broad bandwidth of several gigahertz, we assume that the the noise of the detected signal is dominated by the HEMT thermal noise. We calculate $P_\text{HEMT.0 dBm}$, the signal power at the HEMT while applying \SI{0}{dBm}, by multiplying the noise power of the HEMT (expressed as a noise RMS voltage) with the signal-to-noise ratio (SNR) of the measured signal at \SI{0}{dBm} of applied power. Assuming a 50 $\Omega$ impedance, 
\begin{equation}
\begin{aligned}
    P_\text{HEMT.0 dBm} & = 10\log \left[ \frac{\left(V_\text{Noise RMS}^\text{HEMT}*\text{SNR}_\text{0 dBm}\right)^2}{50/1000}\right]\\
    \\
V_\text{noise RMS}^\text{HEMT} & = \sqrt{P_\text{Noise.HEMT} *50}\\
    \\
P_\text{Noise.HEMT} & = k_{\mathrm{B}} T_{\mathrm{HEMT}} \Delta f\\
\end{aligned} 
\end{equation}

We take noise of the HEMT $P_\text{Noise.HEMT}$ from the noise temperature ($T_{\mathrm{HEMT}}=2$ K) stated in the datasheet of the amplifier. The intermediate frequency bandwidth $\Delta f$ is set when measuring with the VNA and in our case was \SI{5}{kHz}. The SNR is calculated by measuring the magnitude of the S$_{21}$ response of a selected frequency range 50 times and dividing the average signal by the standard deviation. By assuming another \SI{2}{dB} of attenuation between the HEMT and the sample caused by installed circulators, we find the signal power at the sample to be:
\begin{equation}
    P_\text{on\ chip}[\text{dBm}] = 2\text{dB} + P_\text{HEMT.0 dBm} + P_\text{applied}
\end{equation}
The attenuation is given by:
\begin{equation}
    \text{Att}[\text{dB}]=P_\text{on\ chip} - P_\text{applied}
\end{equation}

\begin{figure}[H]
    \centering
    \includegraphics[origin=c, width=\linewidth]{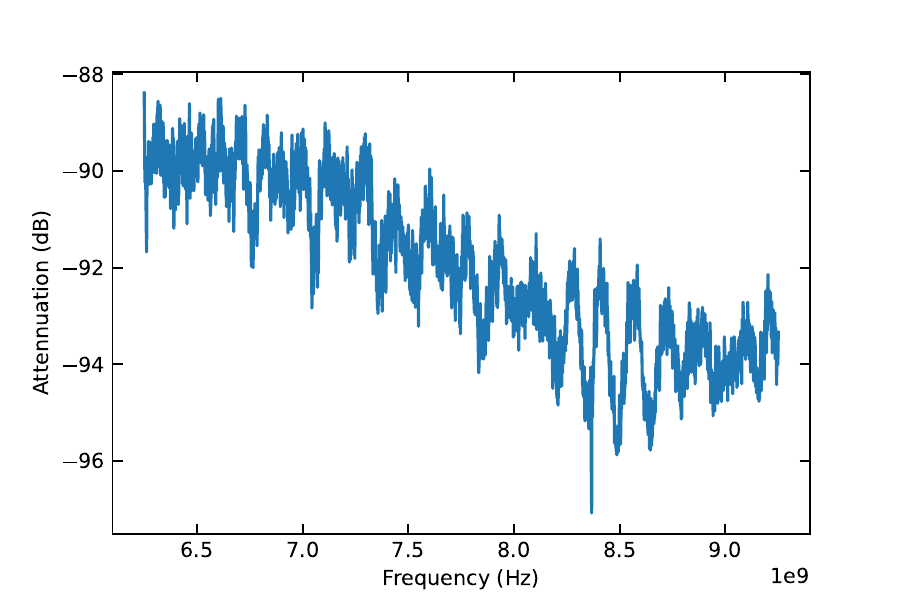}
    \caption{A representative attenuation spectrum over a range that covers the designed frequencies of our resonators. A narrow dip corresponding to a resonance is visible at $\approx8.3$~GHz}
    \label{fig:attenuation}
\end{figure}

The resulting attenuation for one measurement setup is plotted in Fig.~\ref{fig:attenuation}.

\section{Microwave Loss Model}\label{apdx:microwaveLossModel}
From the standard tunneling model~\cite{Phillips1987,Anderson1972}, the temperature and power dependence of the internal quality factor due to TLS loss can be described by~\cite{Gao2008}
\begin{equation}
\label{eq:TLS_simple}
\begin{aligned}
    \frac{1}{Q_{\mathrm{int}}^\mathrm{TLS}} = \delta_{\mathrm{Loss.TLS}} \frac{\tanh \left(\frac{\hbar \omega_r}{2 k_{\mathrm{B}} T}\right)}{\sqrt{1+\frac{n_{\mathrm{ph}}^{\alpha}(P)}{n_{\mathrm{s}}(T)}}} 
\end{aligned}
\end{equation}
where $\delta_{\mathrm{Loss.TLS}}=\sum^N_ip_i\delta_\mathrm{TLS.i}$ is the sum over different lossy regions $i$, each with loss tangent $\delta_\mathrm{TLS.i}$ and participation ratio $p_i$. The participation ratio of a volume $i$ is the fraction of the total electric field strength concentrated in that volume $p_i=\int_{V_i} \frac{\epsilon_i}{2}\left|\mathbf{E}_i(\mathbf{r})\right|^2 / E_{\mathrm{tot}}\ \mathrm{d} \mathbf{r}$~\cite{Wang2015}. $n_\mathrm{ph}(P)$ is the power-dependent intracavity photon number, $\alpha$ is a fitting parameter that is introduced to account for the effect of non-uniform field distributions over the length of the transmission line that makes up the resonator~\cite{Wang2009b,Crowley2023}, $\omega_r=2\pi f_r$ is the angular resonance frequency of the resonator and $n_\mathrm{s}(T)$ is the photon number at which the TLS saturate. 

We cannot quantitatively distinguish between the contributions of different dielectrics to our total loss in this study. To isolate loss contributions from different interfaces (e.g. the substrate-air interface), we would need to vary the cross-sectional geometry of our resonators~\cite{McRae2020}, while to differentiate between loss contributions from different dielectric materials we would need to identify the dielectric species present on each wafer by means of XPS or EELS~\cite{Altoe2022}. Therefore, we only fit the overall sum of dielectric losses $\delta_{\mathrm{Loss.TLS}}$ in our model. This description is still sufficient in our case, as we have qualitative arguments to exclude contributions of different interfaces (see the fourth paragraph of~\autoref{sec:discussion}). 

The saturation photon number $n_\mathrm{s}(T)$ is proportional to $\left(\tau_1(T)\tau_2(T)\right)^{-1}$, the inverse of the averaged relaxation and decoherence times of the TLS ensemble~\cite{vonSchickfus1977,Goetz2016,Crowley2023}. According to the spin-boson model, we expect the relaxation time to follow a thermal distribution $\tau_1 \propto \tanh \left(\frac{\hbar \omega_r}{2 k_{\mathrm{B}} T}\right)$, while we model decoherence from TLS-TLS and TLS-phonon interactions as a simple inverse polynomial relation $\tau_2 \propto T^{-\beta}$~\cite{Gao2008,Burnett2014,Goetz2016,Crowley2023}. This yields the expression $n_{\mathrm{s}}(T)\approx DT^{\beta}\coth\left(\frac{\hbar \omega_r}{2 k_{\mathrm{B}} T}\right)$ which describes the temperature dependence of the saturation photon number. Our TLS model (\autoref{eq:TLS_full}) includes four fit parameters: the exponents $\alpha$ and $\beta$, a proportionality constant of the saturation photon number $D$, and the TLS loss $\delta_{\mathrm{Loss.TLS}}$~\cite{Crowley2023}:
\begin{equation}
\label{eq:TLS_full}
\begin{aligned}
    \frac{1}{Q_{\mathrm{int}}^\mathrm{TLS.full}} = \delta_{\mathrm{Loss.TLS}} \frac{\tanh \left(\frac{\hbar \omega_r}{2 k_{\mathrm{B}} T}\right)}{\sqrt{1+\frac{n_{\mathrm{ph}}^{\alpha}(P)}{DT^{\beta}}\tanh \left(\frac{\hbar \omega_r}{2 k_{\mathrm{B}} T}\right)}}.
\end{aligned}
\end{equation}

To model power-independent but potentially temperature-dependent losses besides TLS loss, we introduce a fifth, temperature-dependent fitting parameter $\delta_\mathrm{PI}$, resulting in~\autoref{eq:tls_fit} as our final loss model.

To apply~\autoref{eq:tls_fit} to our data, we need to determine the photon number in the resonator $n_{\mathrm{ph}}(P)$ as a function of applied microwave power. We can convert microwave power to photon number using the expression~\cite{Wang2009b,Sage2011}
\begin{equation}
\label{eq:photnr}
    n_\mathrm{ph} = \frac{P_\mathrm{in}}{h f_\mathrm{r}^2}.
\end{equation}
where $h$ is the Planck constant. The power in the resonator $P_\mathrm{in}$, also referred to as the `circulating power', can be derived from input-output theory~\cite{Haus1984} to be given as~\cite{Wang2009b,Sage2011}
\begin{equation}
\label{eq:circpower}
    P_\mathrm{in} = P_\mathrm{on\ chip} \frac{Q_\mathrm{int}^2Q_\mathrm{c}}{\pi\left(Q_\mathrm{int}+Q_\mathrm{c}\right)^2}.
\end{equation}
where $P_\mathrm{on\ chip}$ is the applied power after attenuation. In cases where the response of the resonator shows duffing behavior a different formalism for the calculation of the photon number in the resonator is used (see Appendix~\ref{apdx:duffing})

We did not model the effect of dielectric loss on the resonance frequency, because the accuracy of our frequency measurements and fitting procedure was only sufficient to resolve relatively large frequency shifts (on the order of $\Delta f_r/f_0\approx10^{-5}$, likely caused by excess quasiparticles) but was not sufficient to resolve the comparably smaller fractional frequency shifts caused by TLS (which are on the order of $\Delta f_r/f_0\approx10^{-6}$~\cite{Bruno2015,Crowley2023}). 

\section{Fitting procedures}
\label{apdx:fittingProcedures}

\subsection{Fitting S$_{21}$ microwave data}
We analyze the complex S$_{21}$ scattering matrix element of the microwave signal that is sent through the signal chain. From transmission line theory one can derive the S$_{21}$ response of the equivalent lumped LC-circuit for a notch-type configuration~\cite{Pozar2011,Probst2015}
\begin{equation}
    S^{ideal}_{21} = 1-\frac{\left(Q_l /\left|Q_c\right|\right) e^{i \phi}}{1+2 i Q_l\left(f / f_r-1\right)},
\end{equation}
with the loaded quality factor $Q_\mathrm{l}$, the coupling quality factor $Q_\mathrm{c}$, the frequency $f$, and the resonance frequency $f_r$. We include asymmetries in the response due to an impedance mismatch in the phase $e^{i \phi}$ of $Q_\mathrm{c}$ and fit the phase $\phi$~\cite{Probst2015,Khalil2012}. 
We extend the ideal model by adding a complex prefactor to account for an imperfect background with a cable damping $a$, a frequency-dependent cable delay $\tau$, and an initial phase offset $\alpha$~\cite{Probst2015}
\begin{equation}
\label{real_fit}
    S^{real}_{21} = a e^{i \alpha} e^{-2 \pi i f \tau}\left[1-\frac{\left(Q_l /\left|Q_c\right|\right) e^{i \phi}}{1+2 i Q_l\left(f / f_r-1\right)}\right].
\end{equation}
While $Q_\mathrm{c}$ describes energy dissipation into the waveguide through the coupling channel, we are mainly interested in the internal quality factor $Q_\mathrm{int}$, which describes all other sources of loss, including material losses. We can determine $Q_\mathrm{int}$ from the relation $Q_\mathrm{l}^{-1}=Q_\mathrm{int}^{-1}+\Re(Q_\mathrm{c}^{-1}) =Q_\mathrm{int}^{-1}+|Q_\mathrm{c}|^{-1}\cos{\phi}$, following the diameter correction method formalism~\cite{Khalil2012}. By fitting the measured scattering responses with \autoref{real_fit}, we extract the external and internal quality factors and the resonance frequency of each resonator as a function of temperature and applied microwave power.

\begin{figure}
    \centering
    \includegraphics[width = 1\linewidth]{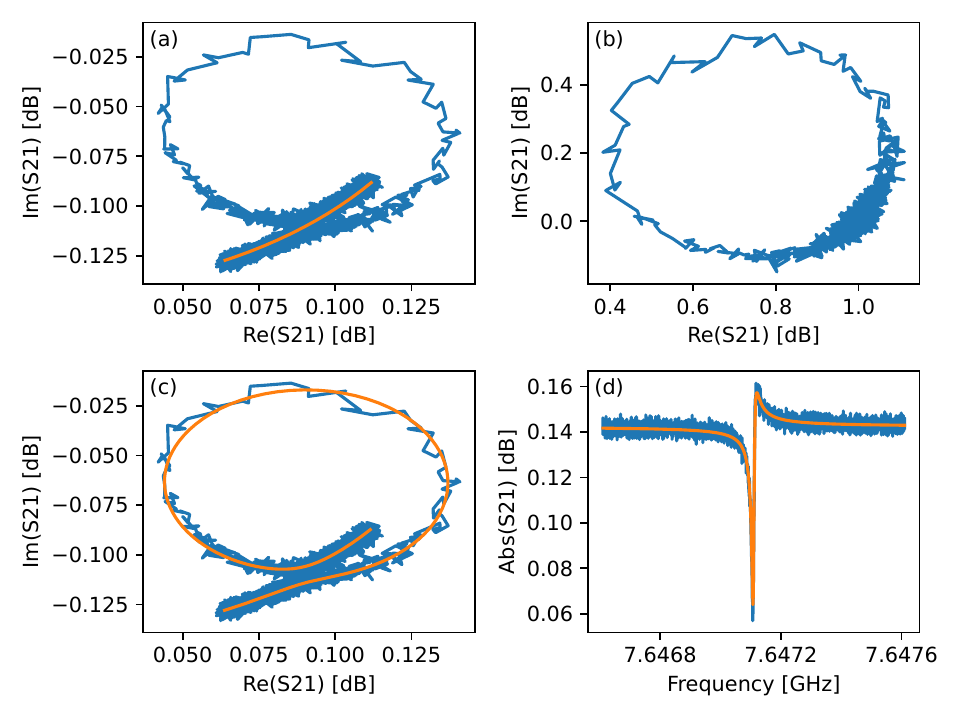}
    \caption{(a) shows the initial background fit of the resonance circle. (b) shows the resonance circle with the background removed. (c) shows the final fit of the resonance circle including the final background refit. (d) shows the final fit of the signal magnitude. The data is taken from one resonator of Wafer \textbf{B} at -50 dBm applied power.}
    \label{fitting_routine}
\end{figure}

We perform a two-step least-squares fit of our data with~\autoref{real_fit}. As a first step, we remove the data around the resonance peak and fit the surrounding background with the expression for the imperfect background in~\autoref{real_fit}. As a second step, we use the parameters of the fitted background as initial guesses and fit the complete model, including a refit of the background parameters. We used a modified version of the stlab~\cite{Stlabgithub} fitting routine which employs the Python library lmfit. This method automatically estimates the standard deviation for all of our fits. The individual steps are shown in~\autoref{fitting_routine}.

\begin{figure}
    \centering
    \subfigure[]{
    \includegraphics[width = 0.9\linewidth]{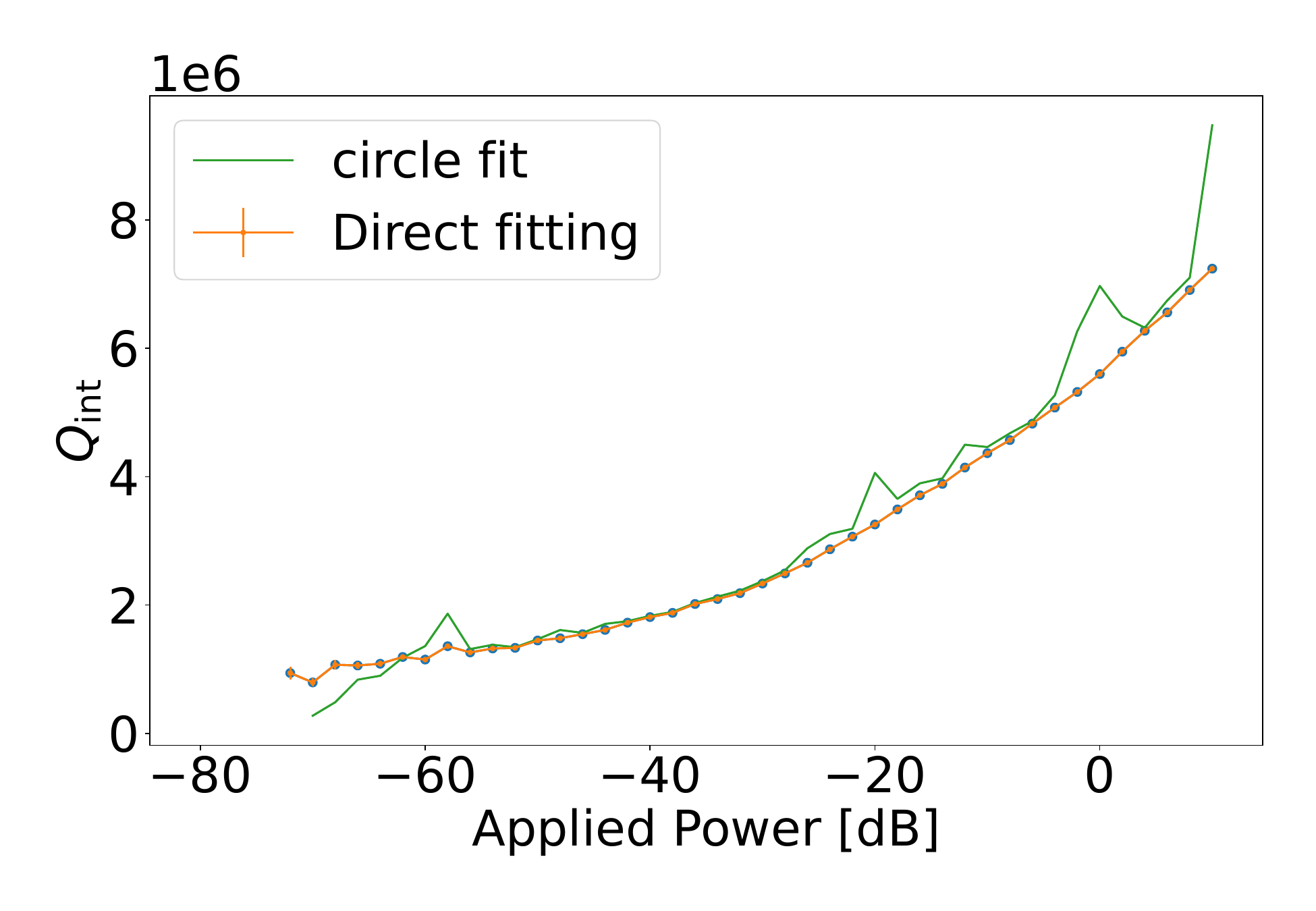}
    \label{fit1}
    }
    \subfigure[]{
    \includegraphics[width = 0.9\linewidth]{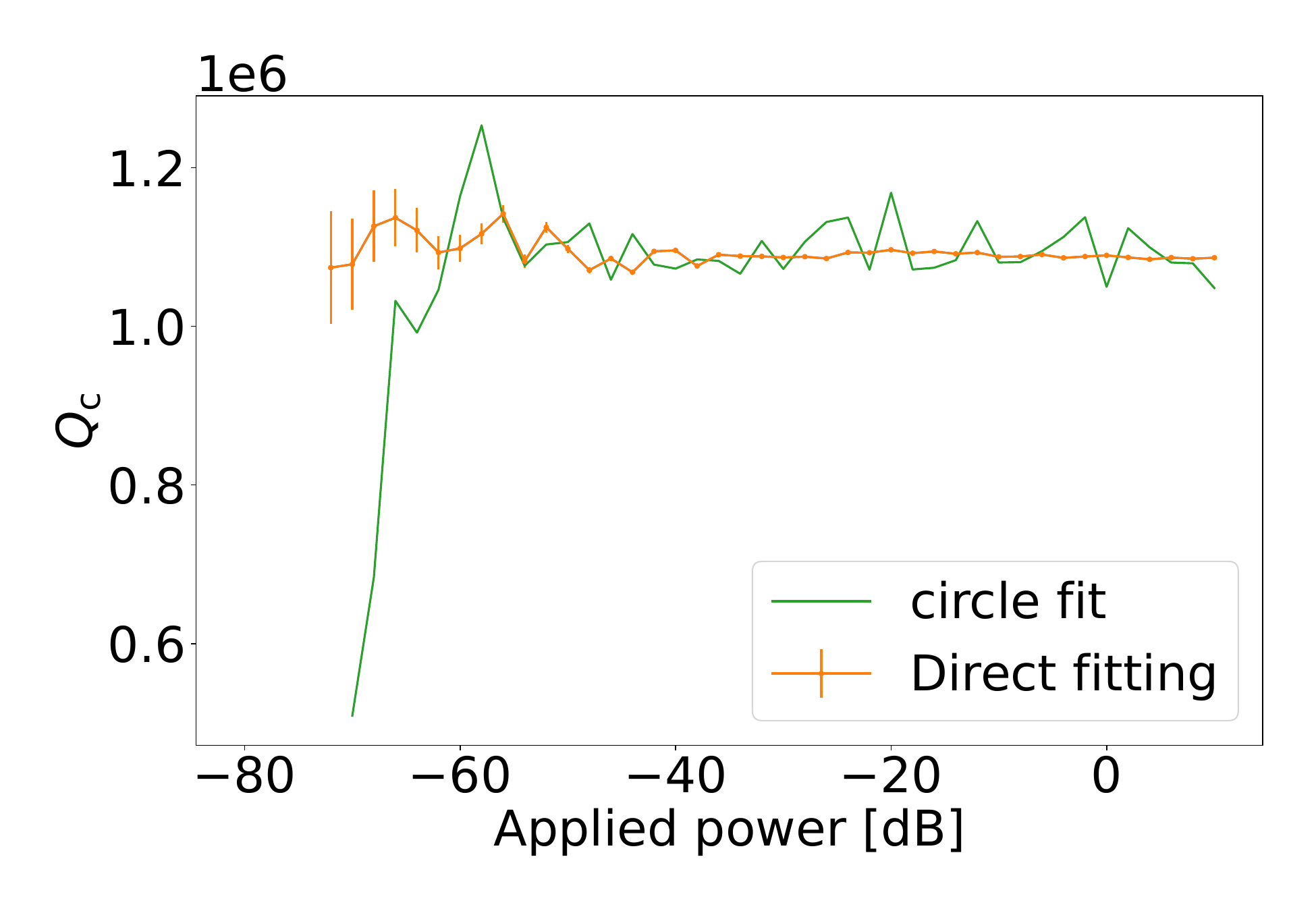}
    \label{fit2}
    }
    \caption{Comparison of the direct fitting routine adapted from~\cite{Stlabgithub} to the fitting routine employed by Probst et al.~\cite{Probst2015}. (a) Fitted internal quality factor versus power for one resonator from Wafer \textbf{B}. (b) Fitted coupling quality factor for the same resonator}
    \label{comp_circ}
\end{figure}

In~\autoref{comp_circ} we show a comparison between the fitted quality factors obtained from the commonly used circle fit method~\cite{Probst2015} and the direct fitting of~\autoref{real_fit} for one representative power trace of one resonator from Wafer \textbf{B}. Both methods fit the same physical model but with different numerical procedures. The obtained values only differ significantly at the lowest applied powers, where the circle fit method yields a strongly divergent coupling quality factor. Since divergent behavior of $Q_\mathrm{c}$ is unphysical, we trust the values obtained by the direct fit more and use the direct fit as our standard fitting procedure.

\label{apdx:duffing}
\begin{figure}
    \centering
    \includegraphics[width = 1\linewidth]{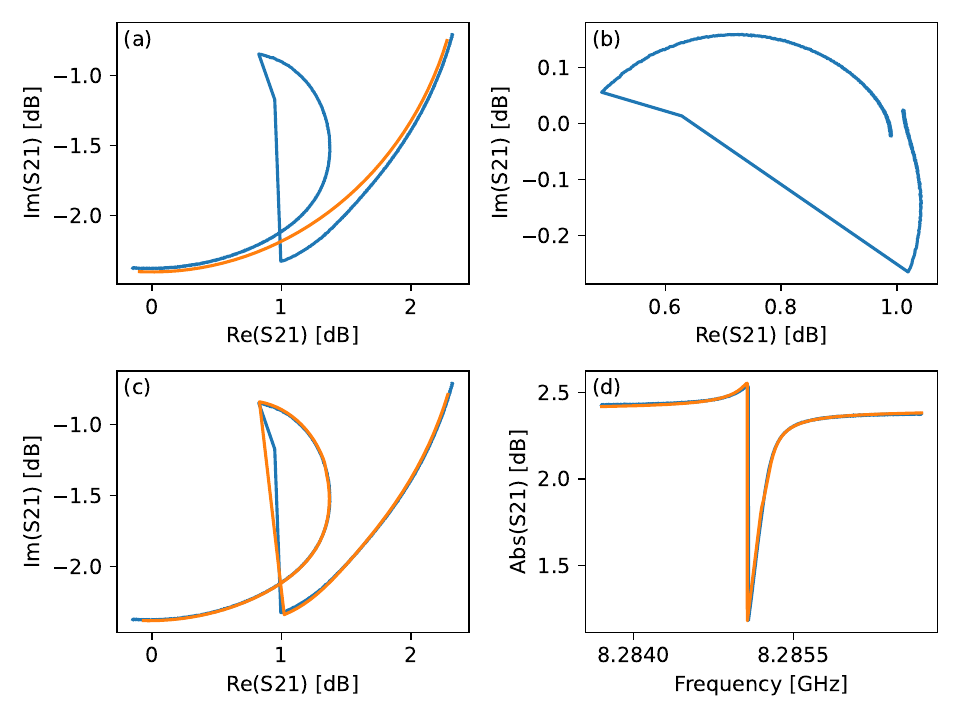}
    \caption{(a)-(d) show the same fitting routine as displayed in~\autoref{fitting_routine} but using the nonlinear model from~\autoref{eq:duffing}. The data is taken from one resonator of Wafer \textbf{C} at -26 dBm applied power.}
    \label{fitting_routine_duffing}
\end{figure}
Some resonators show Duffing-type nonlinear behavior at the highest measurement powers that our standard fit was not able to capture (see~\autoref{fitting_routine_duffing}). The Duffing-type behavior is likely caused by a Kerr nonlinearity and can be modeled as a shift in the resonance frequency of the resonator with microwave power~\cite{Lifshitz2008}. This shift is introduced as a squared term in the equation of motion for the mode amplitude~\cite{Rodrigues2021}:
\begin{equation}
\dot{\alpha}+\left(i\left(\omega_r+\beta|\alpha|^2\right)+\frac{\kappa}{2}\right) \alpha=0,
\end{equation}
where $\alpha$ is the  mode amplitude, $\beta$ the Kerr anharmonicity, $\kappa =\frac{f_r}{Q_\mathrm{l}}$ is the total decay rate. For a derivation of this expression without the anharmonicity term, refer to Sec. 7.2 in~\cite{Haus1984}. 
We can express the scattering parameter in terms of the mode amplitude $\alpha = |\alpha|e^{-i\psi}$ with a phase $\psi$~\cite{Bothner2022}
\begin{equation}
\label{eq:duffing}
\begin{aligned}
    S_{21}^\mathrm{ideal.duffing} & =1+\mathrm{i} \sqrt{\frac{\kappa_{\mathrm{c}}}{2}} \frac{|\alpha|e^{-i\psi}}{\sqrt{\Phi_\mathrm{applied}}}e^{i\phi}\\
    \psi & =\operatorname{atan} 2\left(\frac{2\left(\Delta-\beta |\alpha|^2\right)}{\kappa}\right)
    \end{aligned}
\end{equation}
with $e^{i\phi}$ being introduced to account for impedance mismatches, $\Delta = \omega - \omega_r$ and $\Phi_\mathrm{applied}= \frac{P_\mathrm{applied}}{h f_\mathrm{r}}$ the photon number flux due to the power that is applied to the resonator. From the coupling decay rate $\kappa_\mathrm{c}=\frac{f_r}{|Q_\mathrm{c}|}$ and the total decay rate $\kappa$ we can determine the internal decay rate $\kappa_\mathrm{int}=\frac{f_r}{Q_\mathrm{int}}$ by applying again the diameter correction method $\kappa = \kappa_\mathrm{int} + \kappa_\mathrm{c}\cos{\phi}$. The mode magnitude is given by $|\alpha| = \sqrt{n_\mathrm{ph}}$ where $n_\mathrm{ph}$ can be found as the smallest real positive root of the mode equation rewritten as a third-order polynomial~\cite{Rodrigues2021}
\begin{equation}
\label{eq:photonnumber}
\beta^2 n_{\mathrm{ph}}^3-2 \Delta \beta n_{\mathrm{ph}}^2+\left(\Delta^2+\frac{\kappa^2}{4}\right) n_{\mathrm{ph}}-\kappa_{\mathrm{e}} \Phi_\mathrm{applied}=0 .
\end{equation}

\label{Apdx:Coupling_regimes}
For all methods of fitting the S$_{21}$ microwave data, we observe a strong decrease in fit accuracy when the internal quality factor exceeds the external quality factor. For $Q_\mathrm{int} \gg Q_\mathrm{c}$, the resonator enters the overcoupled regime~\cite{Göppl2008} where coupling losses dominate over internal losses. In this regime the resolution of $Q_\mathrm{int}$ becomes increasingly low and the uncertainty of the fit increases. For this reason, we excluded fitted internal quality factors for which $Q_\mathrm{int} \geq 10\ Q_\mathrm{ext}$. This was not a problem for Wafers \textbf{A} and \textbf{C}, however, one resonator of Wafer \textbf{B} and one resonator of Wafer \textbf{D} could not be fitted for this reason. We note here that measuring in a strongly undercoupled regime $Q_\mathrm{int} \ll Q_\mathrm{c}$ is not ideal either, as high microwave powers are necessary to introduce photons in the resonator. Driving the whole system at high power renders the system more susceptible to heating and other sources of noise.

\begin{figure*}
    \centering
    \includegraphics[width=\textwidth,origin=c]{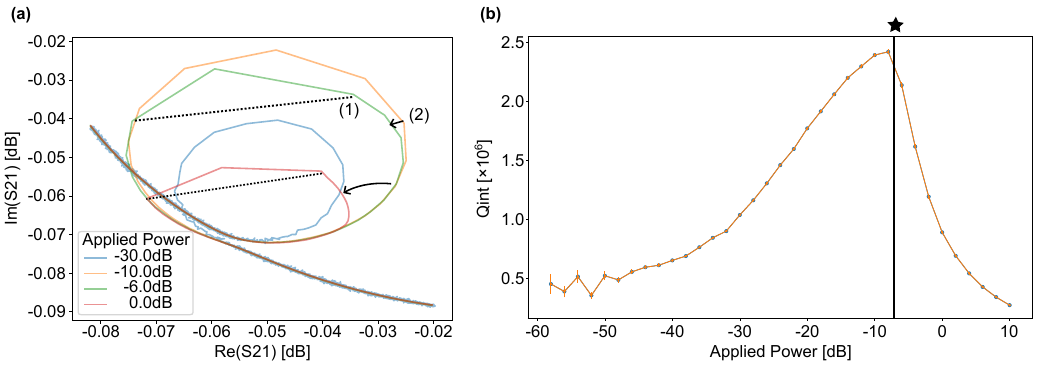}
    \caption{High-power behavior of one resonator from Wafer \textbf{D}. In the raw data (a), the resonance circle starts to feature a jump (1) and caves in towards the jump (2) around -6.0 dB. The abrupt jumps (1) are a sign of duffing behavior and can be fitted, however, our fit can not account for the deformation of type (2). In (b) we show the decrease in internal quality factor at the highest powers calculated by our fitting routine. The star marks the onset of behavior (2).}
    \label{fig:high_power_behavior}
\end{figure*}

At the highest microwave powers that we measured, we observe behavior that is not easily fitted by our standard loss model. The resonance circle starts to appear deformed and/or features jumps due to nonlinear dissipative behavior~\cite{Thomas2020}. If the resonance circle only features a jump, marked as behavior (1) in~\autoref{fig:high_power_behavior}, the internal quality factor can still be extracted with~\autoref{eq:duffing}. If the resonance circle features other distortions, like caving in towards the jump, marked as behavior (2) for -6.0 dB and 0.0 dB in~\autoref{fig:high_power_behavior} (a), the fit will return incorrect results as the numerical method tries to fit a regular circle (or in the case of duffing behavior, a regular circle with a jump to the deformed data). Even though it is expected that the internal quality factor decreases at high power~\cite{deVisser2014}, the high-power internal quality factors shown in~\autoref{fig:high_power_behavior} (b) are likely an underestimation due to the aforementioned fitting errors.

\subsection{Fitting of Loss Models}
As described in~\autoref{apdx:microwaveLossModel}, our loss model includes four fitting parameters accounting for TLS loss and one for temperature-dependent loss. We bound the values of the fitting parameters $\alpha$ and $\beta$ to a reasonable range based on the results of Refs.~\cite{Goetz2016,Crowley2023} for better convergence. A non-linear least-squares fit is performed using the lmfit package in Python. The $Q_\mathrm{int}$ values the model is fitted to are weighted by their corresponding standard deviation. We do not include internal quality factors data points from the highest powers beyond the point where the internal quality factor starts to decrease (marked with a star in~\autoref{fig:high_power_behavior}~(b)), as this behavior is not modeled by~\autoref{eq:tls_fit} and, as we argue in the previous section, the fit uncertainty is high.

\begin{figure*}
    \centering
    \includegraphics[width=0.95\textwidth,origin=c,keepaspectratio]{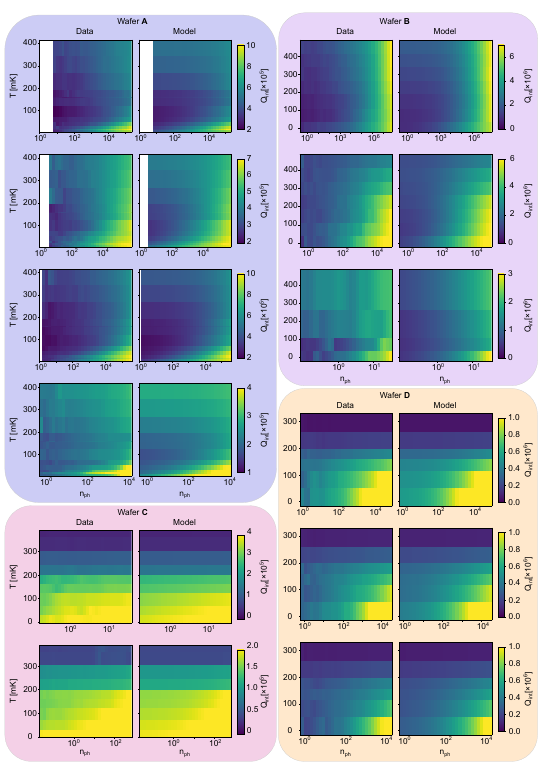}
    \caption{Comparison between data and model for all resonators from included in~\autoref{TLS_figure}~(b)}
    \label{fig:overview_figure}
\end{figure*}

In~\autoref{fig:overview_figure}, we show the results of fitting the power and temperature-dependent data of all resonators included in~\autoref{TLS_figure}~(b). The model was able to reproduce the data well in all cases. 

\section{DC transport characterization}\label{sec:dcTransport}
\begin{figure}[H]
 
    \centering
    \includegraphics[width = 1\linewidth]{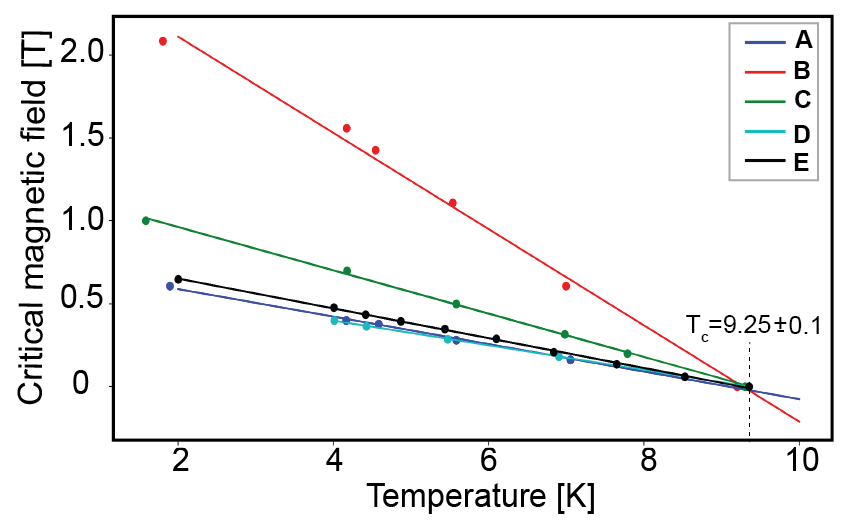}
    
    \caption{DC low-temperature superconducting characteristics of the our films as a function of temperature and magnetic field. The critical temperature $T_C$ is indicated as the point at which the critical magnetic field is zero and is the same for all films.}
    \label{BcTc}
\end{figure}

The full data set from which the values for $\xi_{GL}(0)$ shown in~\autoref{tab:1} are calculated is plotted in~\autoref{BcTc}. Measurements were done at liquid helium temperatures in the Van der Pauw geometry using lock-in amplifiers to measure the relatively low resistances of the films. The bias current was set to 20 $\mu$A in order to get good signal to noise ratios. The critical temperature found for all five wafers was around 9.25$\pm0.1$ K. 

The critical magnetic field at zero K was determined directly from our data by using \cite{Gurevich2003}
\begin{equation}
    B_{c 2}(0)=\left.0.69 T_c \frac{d B_{c 2}}{d T}\right|_{T=T_c}. 
\end{equation}
The Ginzburg-Landau coherence length can then be calculated using 
\begin{equation}
B_{c 2}(T)=\frac{\phi_o}{2 \pi \xi_{G L}^2(T)}.   
\label{Glequation}
\end{equation}
We also determine the mean free path \cite{Mayadas1972}
\begin{equation}
    \rho \ell=3.72 \times 10^{-6} \mu \Omega \mathrm{cm}^2
    \label{mfp}
\end{equation}
which is determined by measuring the wafer's resistivity at 10 K. The resulting values shown in~\autoref{tab:1}. 
\clearpage
\bibliographystyle{apsrev4-2_custom.bst}
\bibliography{papers}

\end{document}